

\let\AMSfonts1   
\magnification1200
\baselineskip=12pt
\hsize=16.5truecm \vsize=23 truecm \voffset=.4truecm
\parskip=14 pt
\def\em{\bf} 
\def\idty{{\leavevmode{\rm 1\mkern -5.4mu I}}}
\def\Ibb #1{ {\rm I\mkern -3.6mu#1}}
\def\ibb #1{\leavevmode\hbox{\kern.3em\vrule
     height 1.5ex depth -.1ex width .2pt\kern-.3em\rm#1}}
\def\Ird{{\hbox{\kern2pt\vbox{\hrule height0pt depth.4pt width5.7pt
    \hbox{\kern-1pt\sevensy\char"36\kern2pt\char"36} \vskip-.2pt
    \hrule height.4pt depth0pt width6pt}}}}
\def\Irs{{\hbox{\kern2pt\vbox{\hrule height0pt depth.34pt width5pt
       \hbox{\kern-1pt\fivesy\char"36\kern1.6pt\char"36} \vskip -.1pt
       \hrule height .34 pt depth 0pt width 5.1 pt}}}}
\def\Ir{{\mathchoice{\Ird}{\Ird}{\Irs}{\Irs} }}
\def\Nl{{\Ibb N}} \def\Cx {{\ibb C}} \def\Rl {{\Ibb R}}
\def\lessblank{\parskip=5pt \abovedisplayskip=2pt
          \belowdisplayskip=2pt }
\def\eproclaim{\par\endgroup\vskip0pt plus100pt\noindent}
\def\proof#1{\par\noindent {\bf Proof #1}\          
         \begingroup\lessblank\parindent=0pt}
\def\QED {\hfill\endgroup\break
     \line{\hfill{\vrule height 1.8ex width 1.8ex }\quad}
      \vskip 0pt plus100pt}

\def\abs #1{{\left\vert#1\right\vert}}
\def\bra #1>{\langle #1\rangle}
\def\bracks #1{\lbrack #1\rbrack}
\def\id{\mathop{\rm id}\nolimits}
\def\ie{i.e.,\ }
\def\ket #1{\vert#1\rangle}

\def\Ketbra #1#2#3{{#1\vert{#2}#1\rangle#1\langle{#3}#1\vert}}
\def\Braket #1#2#3#4{{#1\langle{#2}#1\vert\,{#3}\,#1\vert{#4}#1\rangle}}
\def\norm #1{\left\Vert #1\right\Vert}
\def\Norm #1#2{#1\Vert{#2}#1\Vert}
\def\set #1{\left\lbrace#1\right\rbrace}
\def\Set#1#2#3{\ifx#2:#1\lbrace{#3}#1\rbrace\else
                     {#1\lbrace{#2}\,#1\vert\,{#3}#1\rbrace}\fi}

\def\th {\hbox{${}^{{\rm th}}$}\ }  
\def\tr {\mathop{\rm tr}\nolimits}

\def\limsup{\mathop{\hbox{$\overline{\rm lim}$}}}
\def\phi{\varphi}
\def\epsilon{\varepsilon}
\def\Re{\mathchar"023C\mkern-2mu e}

\ifx\AMSfonts1 \font\frack=eufm10\else
    \let\frack=\bf\fi
\def\qu{q}   
\def\A{{\cal A}} \def\B{{\cal B}} \def\C{{\cal C}}
\def\G{{\cal G}} \def\H{{\cal H}}
\def\K{{\cal K}} \def\Kbar{{\overline{\cal K}}}
\def\E{{\Ibb E}} \def\M{{\cal M}}
\def\ideal{{\cal J}}
\def\SU#1{{\rm SU(#1)}}
\def\SnU#1{{\rm S_{\qu}U(#1)}}
\def\copr{\Delta}
\def\DD#1{{\cal D}^{#1}}
\def\sym{\mathop{\rm sym}\nolimits}
\def\uvec{\vec e}
\def\MF{_{{\rm MF}}}  
\def\XXZ{{{\rm XXZ}}}
\let\y\infty
\def\la{\Lambda}
\def\lap{{\Lambda'}}
\def\lalap{{\Lambda\Lambda'}}
\def\yla{{\infty\Lambda}}   \def\ylap{{\infty\Lambda'}}
\def\pila{{\pi\Lambda}}     \def\piy{{\pi\y}}
\def\dlim{\mkern-2mu\hbox{-}\mkern-5mu\lim} 
\def\jlim{j\dlim}
\def\ilim{i\dlim}
\def\jslim{j^*\dlim}
\def\slim{s\,\dlim}
\def\SS(#1){K({#1})}      
\def\SSG{K_z(\A_\y)}     
\def\sds#1#2{\sigma_{#1},\ldots,\sigma_{#2}}
\def\np{n_+}
\def\nphi(#1,#2){N_{#1}(#2)}  
\def\PHi{{\widehat\Phi}}
\def\Psii{{\widetilde \Psi}}
\def\Phii{{\widetilde \Phi}}
\def\mfac#1!{\prod_{i=1}^{#1}\sqrt{1-\qu^{2i}}}
\def\ppp{\hbox{{\frack p}}}     
\def\hint#1{({#1}\rbrack}       

\def\Fp{F^+}
\def\omegup{\omega^\uparrow}
\def\omegdown{\omega^\downarrow}
\def\bottom{zero-energy\ }
\catcode`@=11
\def\ifundefined#1{\expandafter\ifx\csname
                        \expandafter\eat\string#1\endcsname\relax}
\def\atdef#1{\expandafter\def\csname #1\endcsname}
\def\atedef#1{\expandafter\edef\csname #1\endcsname}
\def\atname#1{\csname #1\endcsname}
\def\ifempty#1{\ifx\@mp#1\@mp}
\def\ifatundef#1#2#3{\expandafter\ifx\csname#1\endcsname\relax
                                  #2\else#3\fi}
\def\eat#1{}
\newcount\refno \refno=1
\def\labref #1 #2 #3\par{\atdef{R@#2}{#1}}
\def\lstref #1 #2 #3\par{\atedef{R@#2}{\number\refno}
                              \advance\refno by1}
\def\txtref #1 #2 #3\par{\atdef{R@#2}{\number\refno
      \global\atedef{R@#2}{\number\refno}\global\advance\refno by1}}
\def\doref  #1 #2 #3\par{{\refno=0
     \vbox {\everyref \item {\reflistitem{\atname{R@#2}}}
            {\d@more#3\more\@ut\par}\par}}\refskip }
\def\d@more #1\more#2\par
   {{#1\more}\ifx\@ut#2\else\d@more#2\par\fi}
\def\@cite #1,#2\@ver
   {\eachcite{#1}\ifx\@ut#2\else,\hskip0pt\@cite#2\@ver\fi}
\def\cite#1{\citeform{\@cite#1,\@ut\@ver}}
\def\eachcite#1{\ifatundef{R@#1}{{\tt#1??}}{\atname{R@#1}}}
\def\defonereftag#1=#2,{\atdef{R@#1}{#2}}
\def\defreftags#1, {\ifx\relax#1\relax \let\next\relax \else
           \expandafter\defonereftag#1,\let\next\defreftags\fi\next }
\def\@utfirst #1,#2\@ver
   {\author#1,\ifx\@ut#2\afteraut\else\@utsecond#2\@ver\fi}
\def\@utsecond #1,#2\@ver
   {\ifx\@ut#2\andone\author#1,\afterauts\else
      ,\author#1,\@utmore#2\@ver\fi}
\def\@utmore #1,#2\@ver
   {\ifx\@ut#2\and\author#1,\afterauts\else
      ,\author#1,\@utmore#2\@ver\fi}
\def\authors#1{\@utfirst#1,\@ut\@ver}
\catcode`@=12
\let\REF\labref
\def\citeform#1{{\bf\lbrack#1\rbrack}}  
\let\reflistitem\citeform               
\let\everyref\relax                     
\let\more\relax                         
\def\refskip{\vskip 10pt plus 2pt}      
\def\colbr{\hskip.3em plus.3em\penalty-100}  
\def\combr{\hskip.3em plus4em\penalty-100}   
\def\refsecpars{\emergencystretch=50 pt      
                 \hyphenpenalty=100}
\def\Bref#1 "#2"#3\more{\authors{#1}:\colbr {\it #2},\combr #3\more}
\def\Gref#1 "#2"#3\more{\authors{#1}\ifempty{#2}\else:\colbr``#2''\fi
                        ,\combr#3\more}
\def\Jref#1 "#2"#3\more{\authors{#1}:\colbr``#2'',\combr \Jn#3\more}
\def\inPr#1 "#2"#3\more{in: \authors{\eds#1}:\colbr
                          ``{\it #2}'',\combr #3\more}
\def\Jn #1 @#2(#3)#4\more{\hbox{\it#1}\ {\bf#2}(#3)#4\more}
\def\author#1. #2,{\hbox{#1.~#2}}            
\def\sameauthor#1{\leavevmode$\underline{\hbox to 25pt{}}$}
\def\and{, and}   \def\andone{ and}          
\def\noinitial#1{\ignorespaces}
\let\afteraut\relax
\let\afterauts\relax
\def\etal{\def\afteraut{, et.al.}\let\afterauts\afteraut
           \let\and,}
\def\eds{\def\afteraut{(ed.)}\def\afterauts{(eds.)}}
\catcode`@=11
\newcount\eqNo \eqNo=0
\def\lasteq{\secNo.\number\eqNo}
\def\deq#1(#2){{\ifempty{#1}\global\advance\eqNo by1
       \edef\n@@{\lasteq}\else\edef\n@@{#1}\fi
       \ifempty{#2}\else\global\atedef{E@#2}{\n@@}\fi\n@@}}
\def\eq#1(#2){\edef\n@@{#1}\ifempty{#2}\else
       \ifatundef{E@#2}{\global\atedef{E@#2}{#1}}%
                       {\edef\n@@{\atname{E@#2}}}\fi
       {\rm($\n@@$)}}
\def\deqno#1(#2){\eqno(\deq#1(#2))}
\def\deqal#1(#2){(\deq#1(#2))}
\def\eqback#1{{(\advance\eqNo by -#1 \lasteq)}}

\def\eqgroup(#1){{\global\advance\eqNo by1
       \edef\n@@{\lasteq}\global\atedef{E@#1}{\n@@}}}
\outer\def\iproclaim#1/#2/#3. {\vskip0pt plus50pt \par\noindent
     {\bf\dpcl#1/#2/ #3.\ }\begingroup \interlinepenalty=250\lessblank\sl}
\newcount\pcNo  \pcNo=0
\def\lastpc{\number\pcNo} 
\def\dpcl#1/#2/{\ifempty{#1}\global\advance\pcNo by1
       \edef\n@@{\lastpc}\else\edef\n@@{#1}\fi
       \ifempty{#2}\else\global\atedef{P@#2}{\n@@}\fi\n@@}
\def\pcl#1/#2/{\edef\n@@{#1}%
       \ifempty{#2}\else
       \ifatundef{P@#2}{\global\atedef{P@#2}{#1}}%
                       {\edef\n@@{\atname{P@#2}}}\fi
       \n@@}
\def\Def#1/#2/{Definition~\pcl#1/#2/}
\def\Thm#1/#2/{Theorem~\pcl#1/#2/}
\def\Lem#1/#2/{Lemma~\pcl#1/#2/}
\def\Prp#1/#2/{Proposition~\pcl#1/#2/}
\def\Cor#1/#2/{Corollary~\pcl#1/#2/}
\def\Exa#1/#2/{Example~\pcl#1/#2/}
\font\sectfont=cmbx10 scaled \magstep2
\def\bgsecti@n #1. #2\e@h{\def\secNo{#1}\eqNo=0}
\def\bgssecti@n#1. #2\e@h{}
\def\secNo{00}
\def\lookahead#1#2{\vskip\z@ plus#1\penalty-250
  \vskip\z@ plus-#1\bigskip\vskip\parskip
  {#2}\nobreak\smallskip\noindent}
\def\secthead#1. #2\e@h{\leftline{\sectfont
                        \ifx\n@#1\n@\else#1.\ \fi#2}}
\def\bgsection#1. #2\par{\bgsecti@n#1. #2\e@h
        \lookahead{.3\vsize}{\secthead#1. #2\e@h}}
\def\bgssection#1. #2\par{\bgssecti@n#1. #2\e@h
        \lookahead{.3\vsize}{\leftline{\bf#1. #2}}}
\def\bgsections#1. #2\bgssection#3. #4\par{%
        \bgsecti@n#1. #2\e@h\bgssecti@n#3. #4\e@h
        \lookahead{.3\vsize}{\vtop{\secthead#1. #2\e@h\vskip10pt
                             \leftline{\bf#3. #4}}}}
\def\Acknow#1\par{\ifx\REF\doref
     \bgsection. Acknowledgements\par#1\refsecpars
     \bgsection. References\par\fi}
\catcode`@=12
\def\class#1 #2*{{#1},}
\overfullrule=0pt
\defreftags AKLT=AKLT, Korepin=AKS, Albev=AHK, Wres=ASW,
AlBaBaBaQ=AB$^{\fam \bffam \tenbf 3}$Q, Alfsen=Alf, Arndt=AHY,
BatBar=BB, BY=BY, Baxter=Bax, BraRo=BR, Drinfeld=Dri, MFD=DW1,
LMD=DW2, Ave=EFS, FCS=FNW1, FCT=FNW2, FCP=FNW3, FCD=FNW4, FQG=FNW5,
Frahm=FYF, FroPfi=FP, Gasp=GR, Gomez=G{\accent 19 o}m, Claudius=Got,
Gui=Gui, Scheunert=HMRS, HSlawny=HS, HM=HM, Jimbo=JMMN, Johnson=Joh,
KLT=KLT, MPG=KSZ, KTasaki=KT, Levy=Lev, math=Mat, Anja=Men,
MezNep=MN, Miek=Mie, Bruno=Na1, Brunogap=Na2, NAG=Nag, Pasq=PS,
Pechersky=Pec, MFRW=RW, Shastri=Sha, Slawny=Sla, Stormer=St\o ,
Delhi=We1, FCL=We2, KAC=We3, CLQ=We4, WoRims=Wo1, WoCMP=Wo2, YY=YY, ,
\line{}
\vskip 2.0cm
\font\BF=cmbx10 scaled \magstep 3

{\BF \baselineskip=25pt
\centerline{Ground states of the infinite }
\centerline{q-deformed Heisenberg ferromagnet}
}
\vskip 1.0cm
\ifx\draft1\centerline{  Version of \today }\fi
\vskip1cm

\centerline{\bf
C.-T. Gottstein$^1$ and
R.F. Werner
\footnote{$^1$}%
{{\sl
FB Physik, Universit\"at Osnabr\"uck, 49069 Osnabr\"uck, Germany
}}%
$^,$\footnote{$^{2}$}
{{ \sl Electronic mail:\quad
   \tt reinwer@dosuni1.rz.Uni-Osnabrueck.DE}}
}
\vskip 1.0cm

{\baselineskip=12pt
\narrower\noindent
{\bf Abstract.}\
We set up a general structure for the analysis of ``frustration-free
ground states'', or ``zero-energy states'', i.e., states minimizing
each term in a lattice interaction individually. The nesting of the
finite volume ground state spaces is described by a generalized
inductive limit of observable algebras. The limit space of this
inductive system has a state space which is canonically isomorphic
(as a compact convex set) to the set of zero-energy states. We show
that for Heisenberg ferromagnets, and for generalized valence bond
solid states, the limit space is an abelian C*-algebra, and all
zero-energy states are translationally invariant or periodic. For
the $q$-deformed spin-$1/2$ Heisenberg ferromagnet in one dimension
(i.e., the XXZ-chain with S$_q$U(2)-invariant boundary conditions)
the limit space is an extension of the non-commutative algebra of
compact operators by two points, corresponding to the ``all spins
up'' and the ``all spins down'' states, respectively. These are the
only translationally invariant zero-energy states. The remaining
ones are parametrized by the density matrices on a Hilbert space,
and converge weakly to the ``all up'' (resp.\ ``all down'') state
for shifts to $-\infty$ (resp.\ $+\infty$).
\par}
\vskip20 pt

\noindent {\bf Mathematics Subject Classification (1991):}
\hfill\break
\class 82B10   Quantum equilibrium statistical mechanics (general)*
\class 82B20   Lattice systems (Ising, dimer, Potts, etc.)*
\class 46A13   Spaces defined by inductive or projective limits (LB, LF,
               etc.)*
\class 46A55   Convex sets in topological linear spaces; Choquet
               theory,*
\hfill\break
\noindent {\bf PACS(1994) Classification:}
\hfill\break
\class 75.10.Jm Quantized spin models*
\class 05.30.-d Quantum statistical mechanics*
\class 03.65.Db Functional analytical methods*
\class 75.25.+z Spin arrangements in magnetically ordered materials*

\vfil\eject
\bgsection 1. Introduction

The determination of all ground states of a given model of
statistical mechanics is usually a very difficult task. Even for a
finite system an explicit characterization of the ground states is
hardly ever possible. Additional problems arise in the passage to
the thermodynamic limit, where, on the one hand, an accidental
ground state degeneracy of the finite volume models can disappear,
or, on the other hand, even if the finite volume models have unique
ground states, some low lying states can converge to additional
ground states \cite{KTasaki}.

In the present paper we study the infinite volume limit of ground
states of quantum models of an especially simple kind: these models
admit states of the infinite system which restrict to ground states
on every finite subregion. For an interaction which is the sum of
translates of a fixed finite range operator, such states not just
minimize the total energy, but even each term in the interaction.
For many (classical or quantum) interactions such states do not
exist, a phenomenon also known as ``frustration''. Therefore we will
call such states ``frustration-free ground states'', or ``\bottom
states''.
In the classical case interactions admitting such ground states are
known as $m$-potentials \cite{HSlawny,Slawny,Miek,Ave}. Often a
lattice interaction which appears to be frustrated allows an
equivalent form which is an $m$-potential  and, in fact, one has to
work hard \cite{Miek} to find examples of ``intrinsically
frustrated'' potentials, for which this is impossible.

For one-dimensional nearest neighbour interactions
$H_2, \widetilde H_2$, equivalence corresponds to perturbations of
the form
$$ \widetilde H_2= H_2+ (X\otimes\idty-\idty\otimes X)
\quad,\deqno(H+bd)$$
with $X$ an arbitrary one-site observable. Clearly, all terms
containing $X$ cancel in the sum over all translates of $\widetilde
H_2$, which represents the formal Hamiltonian of the system.
In particular, the two interactions  generate the same infinite
volume dynamics. That is, for all strictly local observables $A$,
the commutators
$$ \Bigl\lbrack{\sum_{x=-L}^L \widetilde H_{x,x+1},\ A}\Bigr\rbrack
$$
are equal to the same local element ``$\bracks{H,A}$'' for all
sufficiently large $L$, and all perturbations $X$. Hence they have
the same ground states $\omega$, as defined by the property that
$$ \omega\bigl(A^*\bracks{H,A}\bigr)\geq0
\quad,\deqno(gstate)$$
for all strictly local $A$ (see e.g.\ Definition 5.3.18 in
\cite{BraRo}). Note also that $H_2$ and $\widetilde H_2$ have the
same expectation in every translationally invariant state, and since
only such expectations enter the thermodynamic variational
principles, it follows that the two interactions determine the same
thermodynamic functions. For a translationally invariant state
$\omega$, the minimization of $\omega(H_2)$ is in fact equivalent to
\eq(gstate) (see Theorem 6.2.58 in \cite{BraRo}).

The ``\bottom states'' investigated in this paper satisfy a sharper
requirement: the expectation of each translate of $H_2$ is equal to
its smallest possible value, the lowest eigenvalue of $H_2$. These
states are free of ``defects'' \cite{FroPfi}, \ie there are no sites
at which some term in the Hamiltonian can only be minimized by a
global state change. We do {\it not} require translation invariance.
Zero-energy states do satisfy \eq(gstate), and, in fact, proving the
\bottom property is often the best constructive way to show the
ground state property \cite{Shastri}. But in contrast to the ground
state property, the \bottom property now depends on the boundary
term $X$.

An example of this dependence is furnished by the ferromagnetic
spin-$1/2$ XXZ chain, the principal model studied in this paper. The
Hamiltonian is
$$  H_L^\XXZ=\sum_{x=1}^{L-1} \Set\Big:{
                    {-\qu\over2(1 + \qu^2)}
                       \bigl( \sigma_x^1\sigma_{x+1}^1
                   + \sigma_x^2\sigma_{x+1}^2\bigr)
                   + {1\over4}\bigl(\idty-\sigma_x^3
                                           \sigma_{x+1}^3\bigr)}
\quad,\deqno(XXZ)$$
where $\sigma_x^i,\ i=1,2,3$ denotes the Pauli matrices at site $x$,
and $\qu\in(0,1)$ is a parameter. Up to a factor and a constant this
is the Hamiltonian written in \cite{Baxter,YY,Johnson,Korepin} with
anisotropy parameter $\Delta=(\qu+\qu^{-1})/2\,>1$. (Note however,
that relative to some treatments, mainly of the antiferromagnetic
regime of the model \cite{Gomez,Frahm} one has to rotate every other
spin by $\pi$ around the 3-axis, and change the sign of $\Delta$ to
obtain \eq(XXZ)). The ground state vectors of the basic interaction
operator $H_2^\XXZ$ are the product vectors $\ket{++}$ and
$\ket{--}$ (we have chosen constants such that the ground state
energy is $0$). Hence the only way to achieve minimal energy on a
finite chain is to take one of the two product states ``all up'' or
``all down''. Clearly, the corresponding infinite product states,
denoted by $\omegup$ and $\omegdown$, are \bottom states of this
interaction.

It is easy to see that by a suitable perturbation \eq(H+bd) with
$X=\lambda\sigma^1$, one obtains an interaction which does not admit
any frustration-free ground states at all. What is more surprising,
however, is that with another choice of $X$ we can {\it increase}
the number of \bottom states in this model: with
$$ X^\qu={1 - \qu^2\over4(1 + \qu^2)}\ \sigma^3
\deqno(bdX)$$
we get the interaction
$$ H^\qu_2= H^\XXZ_2 + (X^\qu\otimes\idty-\idty\otimes X^\qu)
\quad.\deqno(Hq)$$
This operator is the one-dimensional projection, onto the vector in
$\Cx^2\otimes\Cx^2$ which is invariant under the product
representation $U\otimes U$ of $\qu$-deformed $\SU2$ \cite{WoRims}.
In this sense it is the deformation of the ordinary $\SU2$-invariant
Heisenberg ferromagnetic chain.
The modification \eq(Hq) of the XXZ chain has been considered by
many authors \cite{AlBaBaBaQ,Pasq,MezNep}, although often for
imaginary deformation parameter $\qu$ \cite{Scheunert}, or in the
antiferromagnetic regime \cite{Jimbo}. The finite volume ground
states are considered in \cite{Wres}. However, in that paper no
attempt is made to determine the geometry of the ground state
degeneracy in the infinite system.

It is one of the main aims of this paper to determine the infinite
dimensional manifold of \bottom states of the interaction $H_2^\qu$.
The result is stated in the following Theorem (compare
\Cor17/Statesp/ below, where the the mentioned identification of
states with the density matrices will also be made concrete). Recall
that two states are called {\it quasi-equivalent},  if they are
normal with respect to each other, \ie each one is given by a
density matrix in the representation of the other. Equivalently,
each one is approximated in norm by local perturbations of the
other.

\iproclaim/Tsumm/Theorem.
As a convex set, the set of \bottom states of the
interaction \eq(Hq) is isomorphic to the convex hull of three
quasi-equivalence classes:
\item{(1)} the set consisting only of the ``all spins up'' state
$\omegup$
\item{(2)} the set consisting only of the ``all spins down'' state
$\omegdown$
\item{(3)} a set of ``kink states'', which is isomorphic to the set
of the density matrices on a separable Hilbert space. Each of these
states converges in the w*-topology to $\omegup$ (resp.\
$\omegdown$), when shifted along the chain to right (resp.\ left)
infinity.
\eproclaim

\noindent
The states $\omegup$ and $\omegdown$ are the only translationally
invariant zero-energy states. Since each forms a quasi-equivalence
class by itself, the ground state problem in its GNS representation
is non-degenerate. In contrast, the GNS representation of any kink
state contains all other kink states, so that the ground state
problem is infinitely degenerate. This GNS representation will be
constructed explicitly in Section 4.4. It turns out that in all the
states described in \Thm/Tsumm/ the Hamiltonian in the GNS
representation has a non-zero gap above zero (see Section 4.5, or
\cite{Brunogap}). The gap estimate vanishes precisely for $\qu=1$.
This corresponds to the undeformed Heisenberg ferromagnet, for which
the vanishing of the gap is well-known (arbitrarily low excitations
are given by the so-called magnon states).

Note that all the states described by \Thm/Tsumm/ are also ground
states of the $\XXZ$ interaction. In fact, we get four classes of
ground states, because we can add
{\sl \item{(4)}
the set of ``anti-kink states'', obtained from kink states (3)
by exchanging ``$+$'' and ``$-$''.
\par}\noindent
Of course, the anti-kink states are \bottom states of the
interaction of the form \eq(Hq) with the opposite sign for the
perturbation term. They can also be obtained from the kinks by
left/right inversion of the sites on the chain. This exhausts the
list of ground states arising as \bottom states of perturbations of
the form \eq(Hq). It is not known, however, whether (1),
$\ldots,$(4) is also the complete list of ground states \eq(gstate)
of the $\XXZ$ interaction. In fact, even for the case of the
ordinary Heisenberg ferromagnet (the case $\qu=1$), for which the
\bottom states are well-known (see Section 3.1), this problem seems
to be open (compare the brief discussion at the end of Section 6.2
of \cite{BraRo}).

A classical example with a somewhat similar ground state structure,
two translationally invariant states, and an infinity of
non-invariant kink states, was given by Pechersky \cite{Pechersky}.
An even simpler classical model with this structure is given by the
$q=0$ case of the model studied here: an Ising chain, diagonally
embedded into a quantum chain, in which only the nearest neighbour
configuration ``$+-$'' is forbidden. In this case the ``kink''
states are sharp transitions from ``$+$'' to ``$-$''.

Apart from the treatment of the special model, the aim of this paper
is also the development of some general techniques for the
computation of \bottom state spaces. The basic result (\Thm3/iso/) is
the isomorphism between the set of \bottom states and the state
space of a ``space of zero-energy observables'', which arises in an
inductive limit from the finite volume zero-energy observables. The
inductive limit construction follows closely the usual construction
of the quasi-local observable algebra of the infinite system. It is,
however, not a C*-inductive limit, so that the limit space is not
automatically a C*-algebra.

We illustrate the general construction with two well-known examples.
The first is the Heisenberg ferromagnet (in any dimension, and with
arbitrary couplings). Here the zero-energy algebra is
isomorphic to the algebra of continuous functions on the $2$-sphere.
The points of this sphere correspond precisely to the pure \bottom
states, which are hence characterized by the one direction in space,
along which all spins are directed. The second example is the class
of generalized valence bond solid (GVBS) states \cite{Bruno,FCS,AKLT}.
In this case the zero-energy observables form a finite dimensional
abelian algebra, which implies that every \bottom state has a unique
decomposition into periodic pure states. Moreover, in each extremal
\bottom state the Hamiltonian has a non-zero spectral gap above the
ground state energy. In both these cases the \bottom states retain
some translation symmetry, and hence lie in some simplex of
translation invariant states. The zero-energy algebra must therefore
be commutative. It is perhaps the most interesting feature of the
interaction \eq(Hq) that the set of \bottom states has the structure
of the state space of a {\it non-commutative\/} C*-algebra. This is
only possible due to the lack of translation invariance of the kink
ground states.

Superficially, the $\qu$-deformed Heisenberg ferromagnet has many
features in common with the undeformed ferromagnet ($\qu=1$). For
example, the ground state degeneracy for the model with chain length
$L$ is $L+1$. Moreover, the Hamiltonian is reduced by the
decomposition of the space according to irreducible representations
of deformed $\SU2$ (denoted by $\SnU2$) \cite{WoRims,Drinfeld}, and
these are again in one-to-one correspondence with the
representations of $\SU2$. This would suggest the same ground state
structure for the $\qu$-deformed and undeformed Heisenberg chains.
However, as the Theorem shows, the dimension count is too coarse to
admit such conclusions.

Even though the model we consider is characterized in terms of a
quantum group symmetry, and even though this ``symmetry'' is quite
useful for obtaining an understanding of the ground states in finite
volume, it seems to be of little help in treating the infinite
chain. The fundamental reason for this is that the embeddings used
to define the quasi-local algebra are not consistent with the
quantum group actions defined for each finite volume. In fact, the
quasi-local algebra admits no action of quantum $\SU2$ which
commutes with translations \cite{FQG}. This difficulties are also
seen in the ground state problem: on the intuition derived from
classical symmetries one would expect the ``symmetry'' of the local
Hamiltonians to act on the space of \bottom states. Indeed, this is
the case for the action of the only classical subgroup of $\SnU2$
(the rotations around the $z$-axis), and this fact is crucially used
in our theory to obtain the necessary estimates of operator norms.
However, the spectrum of the generator of this subgroup is the set
of integers with multiplicity one, and this is inconsistent with an
extension of this action to an action of $\SnU2$.

The paper is organized as follows:
In Section 2 we present the general theory of \bottom state problems.
After stating the problem in Section 2.1, we describe, in Section
2.2, the framework for generalized inductive limits of spaces of
observables, and, in particular, the definition of zero-energy
observables of the infinite system.  The connection with Hilbert
space representations of the quasi-local algebra is made in Section
2.3. The general structure is exemplified with the Heisenberg
ferromagnet (Section 3.1) and the generalized valence bond solid
states (Section 3.2). In Section 4 we study the $q$-deformed
Heisenberg ferromagnet. The main result is the identification of the
space of \bottom state observables in Section 4.3 (\Thm16/Limit/).
The Hilbert space representation relevant for the discussion of
\bottom states is described in Section 4.4, and the spectral gap is
estimated in Section 4.5.

\bgsections  2. Zero-energy observables
\bgssection 2.1. Statement of the Problem

In this section we describe abstractly the set of \bottom states of
an infinite quantum lattice system. The basic observable algebra of
the system is thus the quasi-local algebra \cite{BraRo}, which is
constructed as the C*-inductive limit of local algebras, say
$\A_\la$, where $\la$ runs over some collection of finite subregions
of the lattice. For this section, we will only need that the regions
under consideration form a directed set  with respect to inclusion,
so that the notation $\lim_\la x_\la$ for a net $x_\la$ of numbers
makes sense. Readers feeling more comfortable with sequences than
with nets may of course consider all statements along a definite
sequence of increasing regions, and replace the word ``net'' by
``sequence'' throughout.
The algebras $\A_\la$ will be assumed to be finite dimensional
C*-algebras. Typically they are full matrix algebras, \ie $\A_\la$
is the algebra of all linear operators on some finite dimensional
Hilbert space $\H_\la$. We assume that the algebras are embedded
into each other, \ie for $\lap\subset\la$, there is a unit
preserving *-homomorphism
$$ i_\lalap:\A_\lap\to\A_\la
\quad,\deqno(ilalap)$$
satisfying $i_\lalap\circ i_{\lap\la''}=i_{\la\la''}$ whenever
$\la\supset\lap\supset\la''$. For matrix algebras this amounts to
saying that, up to suitable Hilbert space isomorphisms, $\H_\lap$ is
a tensor factor in $\H_\la$, and in the applications we have indeed
that $\H_\la=\H_\lap\otimes\H_{\la\setminus\lap}$.

We briefly recall the construction of the quasi-local algebra,
denoted here by $\A_\y$, since the space of ground state observables
will be defined by a modification of this construction. ``(Strictly)
local'' observables are nets of observables $A_\la\in\A_\la$ with
the property that, for some $\lap$ and all $\la\supset\lap$, we have
$A_\la=i_\lalap(A_\lap)$. Such nets obviously form a *-algebra with
respect to $\la$-wise operations, and since the *-homomorphisms
$i_\lalap$ are automatically isometric, the norm of such nets is
also unambiguously defined as the norm $\norm{A_\la}$, taken for
sufficiently large $\la$. As usual, the C*-inductive limit space
$\A_\y$ of the system $(\A_\la,i_\lalap)$ is now defined as the
completion of the normed space of strictly local observables. The
image of a local observable $A_\la=i_\lalap(A_\lap)$ in $\A_\y$ will
be denoted by $i_\ylap(A_\lap)$. The restrictions of a {\it state}
$\omega$ on $\A_\y$ to the local algebras $\A_\la$ are given by
$\omega_\la=\omega\circ i_\yla$. Conversely, every net of states
$\omega_\la$ on $\A_\la$ satisfying $\omega_\lap=\omega_\la\circ
i_\lalap$ determines a unique state on $\A_\y$. In the sequel, we
will use $\SS(\A)$ to denote the state space of a C*-algebra $\A$.
For reasons which will become clear in the next section we will not
follow the usual practice of suppressing the maps $i_\lalap$ in the
notation.

Suppose now that in each $\A_\la$ an interaction Hamiltonian $H_\la$
is given. We want to study the ground states of these
Hamiltonians. Since $\A_\la$ is finite dimensional this amounts to
the determination of the eigenspace $\G_\la\subset\H_\la$ belonging
to the lowest eigenvalue of $H_\la$. We will denote the projection
onto $\G_\la$ by $g_\la$. Clearly a state $\omega_\la\in\SS(\A_\la)$
is a ground state for $H_\la$, iff $\omega_\la(g_\la)=1$. Is it
possible that all local restriction of a state $\omega\in\SS(\A_\la)$
are ground states in this sense?
This is the basic type of problem addressed in this
article:

\vtop{
\proclaim Main Problem.
Determine all states $\omega$ on the quasi-local algebra $\A_\y$
such that
$$ \omega\Bigl(i_\yla(g_\la)\Bigr)=1
\quad,\quad\hbox{for all }\quad\la\quad.
\deqno(ssg)$$
Such states will be called {\bf \bottom states}, and their set of
will be denoted by $\SSG$. \par}

It is immediately clear that $\SSG$ is a weak*-closed face in
$\SS(\A_\y)$, \ie all convex components of elements of $\SSG$ are
again in $\SSG$. In analogy to ``exposed faces'', \ie the zero sets
of positive continuous affine functionals on a convex set, we call
such a face {\em locally exposed}. In particular, if $\SSG$ consists
of a single state, this state will be called locally exposed
\cite{FCS,FCL}.

What does this mean in typical models? Usually the underlying
lattice then has a {\em translation symmetry}, which is reflected on
the algebraic level by isomorphisms $\tau_x:\A_\la\to\A_{\la+x}$. We
assume $H_\la$ to be translationally invariant with ``free''
boundary conditions:
$$ H_\la=\sum_{{\textstyle{x:\atop \la_0+x\subset\la}}}
          i_{\la,\la_0+x}\ \tau_x(H_0)
\quad,\deqno(hloc)$$
where $\la_0$ is some ``interaction'' region with Hamiltonian
$H_{\la_0}$. (For a nearest neighbour interaction, $\la_0$ consists
of any site together with its neighbours). Now $\omega\in\SSG$
requires, in the special case $\la=\la_0+x$, that
$$\omega(i_{\y,\la_0+x}\tau_x(H_0))=h_0
\quad,\quad\hbox{for all $x$, }\deqno(locmin)$$
where $h_0$ is the lowest eigenvalue of $H_0$. Conversely,
\eq(locmin) implies that {\it every term} in the $\omega$-expectation
of \eq(hloc) is equal to its lowest possible value $h_0$. Hence
$\omega_\la$ is a lowest energy state for every $H_\la$, \ie
$\omega\in\SSG$.

For generic interactions one typically finds $\SSG=\emptyset$,
\ie most interactions are ``frustrated''. In the quantum case
this even happens in one dimension \cite{FCL}. A fundamental example
is the spin-$1/2$ nearest neighbour Heisenberg antiferromagnet. In
that model $\omega\in\SSG$ requires the state to be supported by the
antisymmetric subspace for any pair of nearest neighbours, and hence
on the antisymmetric subspace of every $\H_\la$, which vanishes
whenever $\la$ contains three or more sites. Hence $\SSG=\emptyset$.
It is not known whether in this case there is an equivalent finite
range interaction allowing some \bottom state. On the other hand,
there are some interesting models with a non-trivial $\SSG$. Three
paradigms are the Heisenberg ferromagnet on an arbitrary lattice
(see Section 3.1), the VBS ground states studied in
\cite{AKLT,FCS,FCP,Bruno,BY} (see Section 3.2), and the deformed
Heisenberg chain studied in Section 4 of this paper. The \bottom
state spaces $\SSG$ are strikingly different in these three cases.

We close this section with the verification of the claim made in the
introduction that \bottom states are ground states in the sense of
the standard definition \eq(gstate). This definition would require
of a state $\omega$, in the notation adopted here, that
$$ \lim_\la \omega\bigl(X^*\bracks{i_\yla(H_\la),\,X}\bigr) \geq0
\quad,\deqno(gstate:i)$$
for all strictly local elements $X=i_\yla(X_\la)\in\A_\y$. But if
$\omega\in\SSG$, we get
$$ \omega\bigl(X^*\bracks{i_\yla(H_\la),\,X}\bigr)
        =\omega\bigl(X^*i_\yla(H_\la)\,X\bigr)-\omega(X^*X)h_\la
        \geq0
\quad,$$
where $h_\la$ is the smallest eigenvalue of $H_\la$. Hence $\omega$
is a ground state. The converse certainly fails in general, since
$\SSG$ may be empty. Even if $\SSG\not=\emptyset$, however, it is
highly unclear which additional conditions make it true.

\bgssection 2.2. The Inductive Limit

For a finite system $\A_\la$ the space of ground states has a very
simple structure: it is simply the state space of the algebra of
operator on the lowest eigenvalue eigenspace $\G_\la\subset\H_\la$.
It is therefore suggestive to define the space of {\em zero-energy
observables} for the finite system as
$$ \B_\la=\B(\G_\la)=g_\la\A_\la g_\la
\quad,\deqno(Bla)$$
where $\B(\H)$ denotes the space of bounded operators on a Hilbert
space $\H$, and $g_\la$ is the projection from $\H_\la$ onto
$\G_\la$ as in the previous subsection. Our task in this subsection
is to define an analogue of $\B_\la$ for the infinite system, whose
state space is canonically isomorphic to $\SSG$. Since there is no
analogue of the projection $g_\la$ in the quasi-local algebra
$\A_\y$, it is clear that equation \eq(Bla) is not suitable for this
purpose.

It is useful to go back to the definition of the quasi-local algebra
as an inductive limit. As described above, with this definition a
quasi-local observable $A_\y\in\A_\y$ becomes an equivalence class
of Cauchy nets whose members are again nets. This sounds rather
involved, but causes no technical complications, because the
strictly local observables $A_\la=i_\lalap A_\lap$ are so simple
that the $\la$-dependence is often suppressed ``by canonical
identification''. This would not be permissible if the embedding
maps $i_\lalap$ were not isometric, or, even more generally, if the
inductive limit relation $i_\lalap\circ i_{\lap\la''}=i_{\la\la''}$
were only satisfied approximately. Such ``approximate inductive
limits'' have become a useful tool in a variety of contexts, e.g. in
the statistical mechanics of mean-field systems \cite{MFD,LMD}, as a
general framework for various thermodynamic limits for intensive
observables \cite{KAC}, or in the formulation of classical
($\hbar\to0$) limit of quantum mechanics \cite{CLQ}. The definition
of the space of zero-energy observables for an infinite system will
be yet another application. In all these cases it turns out to be
convenient to reduce the implicit double sequence construction of
the quasi-local algebra to a construction involving only nets
indexed by $\la$.

We will now briefly review the basic idea of approximate inductive
limits, referring to \cite{KAC,CLQ} for more details and proofs.
So let $(\B_\la,j_\lalap)$ be a system consisting of normed spaces
$\B_\la$ indexed by the elements $\la$ of some directed set, and
contractive maps $j_\lalap:\B_\lap\to\B_\la$. In this situation we
will call a net $B_\la\in\B_\la$ {\em j-convergent}, if
$$ \lim_\lap\limsup_\la \norm{B_\la-j_\lalap(B_\lap)}=0
\quad.\deqno(jconv)$$
In particular, {\em null nets} with $\lim_\la\norm{B_\la}=0$ are
convergent. We will say that two convergent nets ``have the same
limit'', if they differ by a null net. Hence we  define the {\em
limit space} of the system $(\B_\la,j_\lalap)$ as the space of
$j$-convergent nets modulo the space of null nets. This space will
be denoted by $\B_\y$. The {\em limit} of a convergent net $B_\la$
is the class of the net in this quotient, and will be denoted by
$B_\y$, or more explicitly, by $\jlim_\la B_\la\in\B_\y$. One easily
checks that, for $j$-convergent nets, the net of norms is also
convergent, so $\B_\y$ becomes a normed space with
$$ \norm{\jlim_\la B_\la}:=\lim_\la \norm{B_\la}
\quad.\deqno(norm)$$
$\B_\y$ is always complete \cite{KAC}.
Note that if all $\B_\la$ are the same normed space, and all
$j_\lalap$ are the identity operator on this space, the
$j$-convergent nets are just the Cauchy nets, and $\B_\y$ is just
the completion of the given normed space.

Nets $B_\la$ with the property that $B_\la=j_\lalap B_\lap$, for
some $\lap$ and all $\la\supset\lap$, are called {\em basic nets}.
We will always assume that such nets are $j$-convergent, which
expresses the asymptotic transitivity of the comparison furnished by
the maps $j_\lalap$. This condition will be trivially satisfied in
this paper, since we will always have
$j_\lalap\circ j_{\lap\la''}=j_{\la\la''}$ for
$\la\supset\lap\supset\la''$ (compare \eq2.16(chain)). We can then
define maps $j_\ylap:\B_\lap\to\B_\y$ by
$$ j_\ylap(B)=\jlim_\la j_\lalap(B)
\quad,\quad\hbox{for}\quad B\in\B_\lap
\quad.\deqno()$$
It is easy to see that the elements of the form $j_\ylap B$ are norm
dense in $\B_\y$. The basic nets are also dense as a space of nets:
a net $B_\la$ is convergent iff for any $\epsilon>0$ there is a
basic net $B^\epsilon$ such that
$\limsup_\la\norm{B_\la-B_\la^\epsilon}\leq\epsilon$.

In this paper we are also concerned with the limit of states, \ie
with positive normalized functionals. In order that positivity and
normalization make sense in $\B_\y$, we will assume that each
$\B_\la$ is an order unit space \cite{NAG}, e.g.\ a C*-algebra, and
that the $j_\lalap$ preserve both the orderings and the unit
elements, in the sense that $j_\lalap(\idty_\lap)=\idty_\la$, and
that $A\geq0$ implies $j_\lalap(A)\geq0$. This implies that $\B_\y$
has an ordering, for which the positive cone consists of the limits
of all convergent sequences of positive elements, and a unit, namely
$\idty_\y=\jlim_\la\idty_\la$. It can be shown \cite{KAC} that
thereby $\B_\y$ becomes an order unit space, so that we can define
the state space of $\B_\y$ as
$$ \SS(\B_\y):=\Set\big{\omega:\B_\y\to\Cx}
                   {\ \omega\,\hbox{linear},\
                    A\geq0\Rightarrow\omega(A)\geq0,\
                    \omega(\idty_\y)=1}
\quad.\deqno(ssBy)$$
Associated with the definition of $j$-convergent nets (which is a
convergence in norm) there is a notion of weak convergence of
states: we say that a net $\omega_\la\in\SS(\A_\la)$ is
{\em j*-convergent}, if, for any $j$-convergent net
$A_\la\in\A_\la$, the sequence of numbers $\omega_\la(A_\la)$ is
convergent. It is easy to see that in this case a state
$\omega_\y\in\SS(\B_\y)$ is defined by the formula
$$ \omega_\y\bigl(\jlim_\la A_\la\bigr)
     =\lim_\la\omega_\la(A_\la)
\quad.\deqno(jslim)$$
Every state $\omega\in\SS(\B_\y)$ is a $j^*$-limit of such a net of
states, namely of $\omega_\la=\omega\circ j_\yla$.

After this excursion to generalized inductive limits we come back to
the problem of defining zero-energy observables for an infinite
system. It is clear that we must make some assumptions about the
subspaces  $\G_\la\subset\H_\la$ or, equivalently, about the
projections $g_\la\in\A_\la$. Our {\em standing assumptions} will be
$$\eqalignno{
    g_\la&\neq0
\hskip44pt,\quad\hbox{for all $\la$, and}
&\deqal(g>0)\cr
    g_\la&\leq i_\lalap(g_\lap)
\quad,\quad\hbox{for all $\la\supset\lap$.}
&\deqal(g<g)\cr}
$$
These are dictated by the intention to study situations with
$\SSG\neq\emptyset$: the support projections $\tilde g_\la^\omega$
of the restrictions $\omega_\la=\omega\circ i_\lalap$ of any state
$\omega\in\SS(\A_\y)$ automatically satisfy these assumptions.
Moreover, if $\omega\in\SSG$, we get $\tilde g_\la^\omega\leq
g_\la$. Hence even if the $g_\la$ do not satisfy \eq(g<g) initially,
we can pass to smaller projections satisfying both assumptions. It
is also clear that the two assumptions suffice to guarantee
$\SSG\neq\emptyset$: because of \eq(g>0), we can find states with
$\omega_\la(g_\la)=1$. Then by compactness we can find a
weak*-cluster point $\omega_*\in\SS(\A_\y)$ of suitable extensions
of these states. From \eq(g<g) it then follows that
$\omega_*\in\SSG$.

We will only be interested in systems of projections which are
{\em non-trivial} in the sense that the net $g_\la$ of projections
is {\it not} quasi-local. In fact, if $g_\la$ were $i$-convergent,
$i_\yla g_\la$ would be a norm convergent decreasing sequence of
projections, which must be eventually constant, because the norm
difference of commuting projections is either $1$ or $0$. Hence
$g_\la$ would even be strictly local. This is impossible in a
statistical mechanics model, since increasing the region always
introduces more terms in the Hamiltonian to be minimized and hence
more constraints on the state.

We choose the algebras $\B_\la$ as defined in equation \eq(Bla). The
unit in $\B_\la$ is the projection $g_\la\in\A_\la$, and for this
unit to be different from $0$, condition \eq(g>0) must be satisfied.
Then, for each $\la$, the map
$$\eqalign{
    r_\la:\A_\la&\to\B_\la    \cr
        r_\la(A)&:= g_\la A g_\la
\quad,\quad\hbox{for $A\in\A_\la$.}
\cr}\deqno(rla)$$
is positive, unit preserving, and surjective. Furthermore, we
define, for $\la\supset\lap$:
$$\eqalign{
    j_\lalap:\B_\lap&\to\B_\la  \cr
  j_\lalap(r_\lap A)&:=r_\la(i_\lalap(A) )
\quad,\quad\hbox{for $A_\lap\in\A_\lap$.}}
\deqno(jlalap)$$
This is well-defined since, by condition \eq(g<g), $r_\lap(A)=0$
implies
$$\eqalign{
   r_\la(i_\lalap(A) )
   &=g_\la i_\lalap(A)g_\la
    =g_\la i_\lalap(g_\lap)\ i_\lalap(A)\ i_\lalap(g_\lap)g_\la  \cr
   &=g_\la i_\lalap(g_\lap A g_\lap)g_\la
    =g_\la i_\lalap(r_\lap(A))g_\la
    =0
\quad.\cr}$$

Let $\la\supset\lap\supset\la''$. Then $j_\lalap\circ j_{\lap\la''}$
obviously satisfies the defining equation \eq(jlalap) for
$j_{\la\la''}$. Hence we have
$$ j_\lalap\circ j_{\lap\la''}=j_{\la\la''}
\quad.\deqno(chain)$$

Suppose now that $A_\la$ is an $i$-convergent net for the system
$(\A_\la,i_\lalap)$ defining the quasi-local algebra, and consider
the net $B_\la:=r_\la(A_\la)\in\B_\la$. Then
$$  \norm{B_\la-j_\lalap(B_\lap)}
     =\norm{r_\la\bigl(A_\la-i_\lalap(A_\lap)\bigr)}
     \leq\norm{A_\la-i_\lalap(A_\lap)}
$$
goes to zero in just the way required for $j$-convergence of $B_\la$.
The limit $B_\y$ is not changed, if $A_\la$ is modified by a null
net. Hence there is a well-defined operator
$$\eqalign{
   r_\y:\A_\y&\to\B_\y
\qquad\hbox{with}\cr
   r_\y(\ilim_\la A_\la)&= \jlim_\la r_\la A_\la
\quad,}
\deqno(ry)$$
for any $i$-convergent net $A_\la\in\A_\la$. The surjectivity of the
finite volume maps $r_\la$ also goes to the limit, as the following
Lemma shows.

\iproclaim/ryonto/Lemma.
$r_\y$ is surjective, and maps the unit sphere of $\A_\y$ onto a
dense subset of the unit sphere of $\B_\y$.
\eproclaim

\proof:
Consider $B_\y=\jlim_\la B_\la$ with $\norm{B_\y}\leq1$, and fix
some summable sequence $\epsilon_n$. Then we can find a sequence of
regions $\cdots\la_n\subset\la'_n\subset\la_{n+1}\cdots$ such that
$$ \norm{B_\la-j_{\la\la_n}(B_{\la_n})}\leq\epsilon_n
$$
for all $\la\supset\la'_n$. Considering $B_{\la_n}$ as an element of
$\A_{\la_n}$, we can define
$$  A_\la =i_{\la\la_0}(B_{\la_0} +
            \sum_{{n\geq1 \atop \la'_n\subset\la}}
           i_{\la\la_n}\bigl(B_{\la_n}
                -j_{\la_n\la_{n-1}}(B_{\la_{n-1}})\bigr)
\quad.$$
This sum converges in norm, uniformly in $\la$, and we have
$\norm{A_\la}\leq\norm{B_\y}+2\epsilon_0+\sum_{n\geq1}\epsilon_n$.
Moreover, $A$ is $i$-convergent, and $r_\la(A_\la)$ is a telescoping
series evaluating to $B_{\la_n}$ for the largest $n$ such that
$\la'_n\subset\la$. Hence $\norm{r_\la A_\la-B_\la}\to0$, and
$r_\y A_\y=B_\y$. Moreover, if $\norm{B_\y}<1$, we can choose the
$\epsilon_i$ sufficiently small to make $\norm{A_\y}\leq1$.
\QED

The adjoint $r_\y^*$, which is a weak*-continuous map from the dual
$\B_\y^*$ of $\B_\y$ to $\A_\y^*$, is therefore an isometric map,
and is precisely the desired isomorphism between the \bottom states
and the states on $\B_\y$:

\iproclaim/iso/Theorem.
The map $r_\y^*:\SS(\B_\y)\to\SSG\subset\SS(\A_\y)$ is an
isomorphism of compact convex sets, where all state spaces are
equipped with the weak*-topology.
\eproclaim

\proof:
We will explicitly construct the inverse of $r_\la^*$. So let
$\eta\in\SSG$. Since $\B_\la=g_\la\A_\la g_\la$ can be considered as a
subspace of $\A_\la$, we can evaluate the restriction
$\eta\circ i_\yla$ on $\B_\la$.
We define
$$\eqalign{
   s_\la:\SSG&\to\SS(\B_\la) \cr
   s_\la(\eta)(r_\la A)&=\eta\circ i_\yla(A)
\quad,\quad\hbox{for all $A\in\A_\la$.}
}$$

We have to verify first that $s_\la(\eta)$ is a well-defined state.
Because $\eta\circ i_\yla(g_\la)=1$, we have that the right hand side
$\eta\circ i_\yla(A)
   =\eta(i_\yla(g_\la Ag_\la))
   =\eta\circ(i_\yla(r_\la(A)))$
indeed depends only on $r_\la(A)$. Positivity is obvious, and
normalization follows by putting $A=\idty$.

Next, since
$$\eqalign{
  s_\la(\eta)\circ j_\lalap (r_\lap(A))
   &=s_\la(\eta)\circ r_\la\circ i_\lalap(A)
    =\eta\circ i_\yla\circ i_\lalap(A)   \cr
   &=\eta\circ i_\ylap(A)
    =s_\lap(\eta)(r_\lap(A))
\quad,}$$
we have $s_\la(\eta)\circ j_\lalap=s_\lap(\eta)$. This readily implies
that
$$ s_\y(\eta):=\jslim_\la s_\la(\eta)$$
exists, and is a state on $\B_\y$.

We claim that $s_\y$ is continuous for the weak*-topologies. Let
$\eta^\alpha$ is a weak*-convergent net in $\SSG$, which is to say
that, for all $A\in\A_\y$, the net $\eta^\alpha(A)$ is convergent. We
have to show that, similarly, $s_\y(\eta)(B)$ converges for
$B\in\B_\y$. It suffices to verify this for the norm dense subset of
elements of the form $B=j_\ylap(r_\lap A)$ with $A\in\A_\lap$ for some
$\lap$. But for these we have
$$\eqalign{
   s_\y(\eta^\alpha)(B)
     &=\lim_\la s_\la(\eta^\alpha) (j_\lalap\circ r_\lap A) \cr
     &=\lim_\la s_\la(\eta^\alpha) (r_\la\circ i_\lalap A) \cr
     &=\lim_\la \eta^\alpha (i_\lalap A)
\quad,\cr}$$
which is convergent by assumption.

It remains to be proven that $s_\y$ and $r_\y^*$ are inverses of each
other. First, let $\eta\in\SSG$. Then, for $A\in\A_\la$,
$$\eqalign{
   r_\y^*\circ s_\y(\eta)\bigl(i_\yla(A)\bigr)
     &=s_\y(\eta)(r_\y\circ i_\yla(A)))
      =\lim_\la s_\la(\eta)(r_\la(A))      \cr
     &=\eta(i_\yla(A))
\quad.}$$
Hence $r_\y^*s_\y(\eta)=\eta$ on the dense set of elements
$i_\yla(A)$, and hence everywhere. This proves
$r_\y^*\circ s_\y=\id_{\SSG}$.

Conversely, let $\omega\in\SS(\B_\y)$, and $A\in\A_\la$. Then
$$\eqalign{
   s_\y\circ r_\y^*(\omega)\bigl(j_\yla\circ r_\la(A)\bigr)
     &=s_\la\circ r_\y^*(\omega)\bigl(r_\la(A)\bigr)
      =r_\y^*(\omega)\bigl(i_\yla(A)\bigr) \cr
     &=\omega\bigl(r_\y\circ i_\yla(A)\bigr)
      =\omega\bigl(j_\yla\circ r_\la(A)\bigr)
\quad.\cr}$$
Hence $s_\y\circ r_\y^*\omega=\omega$  on a norm dense subset of
$\B_\y$ and, consequently, $s_\y\circ r_\y^*=\id_{\SS(\B_\y)}$.
\QED

The upshot of this Theorem is not so much that $\SSG$ is identified
as the state space of some order unit space $\B_\y$. In fact, that is
true of any compact convex set \cite{Alfsen}. It is rather that this
space is obtained by a direct construction, which is based on the
asymptotic behaviour of the finite volume ground state spaces.
One can use this to transfer some of the properties of the finite
volume spaces to the limit more easily. A special role in this
regard is played by the algebraic product. Although each $\B_\la$ is
a C*-algebra, and the embeddings $j_\lalap$ are completely positive
and unit preserving, they are not homomorphisms. Therefore, the
limit space does not automatically inherit the product operation.
However, it is true in all the examples below that $j_\lalap$ is
asymptotically a homomorphism, in a sufficiently strong sense to
make $\B_\y$ into a C*-algebra, as well. We state this special
situation in the following Definition for later reference.

\iproclaim/D.prod/Definition.
The inductive system $(\B_\la,j_\lalap)$ is said to have the {\bf
product property}, if, for any two $j$-convergent nets
$A_\la,B_\la$, the net defined by $C_\la=A_\la B_\la$ for every $\la$,
is also $j$-convergent.
\eproclaim

If the product property holds, we can immediately define a product
on $\B_\y$ by
$$  A_\y B_\y=C_\y
\quad,\deqno(By-prod)$$
and it is clear that $\B_\y$ thereby becomes a C*-algebra. Of course,
this makes the determination of the state space much easier, since
much more is known about C*-algebras than about general order unit
Banach spaces. We have not found a general way of proving the
product property. At least the proofs in the three cases considered
below are quite different, and are based on specific properties of
each of the models. For the undeformed ferromagnet the proof also
yields the commutativity of the product, and consequently $\SSG$  is
a simplex. On the other hand, in the deformed case considered in
Section 4, the product is well-defined, but non-commutative. In
other words, there are not only convex combinations, but ``coherent
superpositions'' of \bottom states, as well.

\noindent{\parskip=5pt
With this Theorem we arrive at the following procedure for computing
$\SSG$:
\item{(1)}
determine the local ground state spaces $\G_\la$ from the given
Hamiltonians.
\item{(2)}
Compute the inductive limit space $\B_\y$.
Decide the product property.
\item{(3)}
Determine the state space of $\B_\y$.}

\noindent
In the cases considered below (1) is fairly easy. Step (2) is
usually the most difficult part. Since the spaces $\B_\y$ arising in
these examples are quite simple, (3) is trivial. The main work thus
goes into (2).

\bgssection 2.3. Hilbert space representations

For the model considered in Section 4 it is easy to find a
representation of the quasi-local algebra with respect to which many
of the states in $\SSG$ are obviously normal. (Recall that a state
$\omega$ on a C*-algebra $\C$ is called normal in a representation
$\pi:\C\to\B(\H_\pi)$, if there is a trace class operator
$D_\omega$ on $\H_\pi$ such that
$\omega(C)=\tr\bigl(D_\omega \pi(C)\bigr)$).
In that example it is also true, but much more difficult to show
that (with two exceptions) {\it all} states $\omega\in\SSG$ are
normal in this representation. Working inside just one
representation, such questions are impossible to decide, and it was
precisely for obtaining such complete characterizations of
$\SSG$ that the inductive limit scheme of Section 2.2 was set up.
Nevertheless, representations are a useful tool. The aim of this
section is to describe briefly how representations of $\A_\y$
generate representations of $\B_\y$, and under which circumstances
such representations are faithful.

Let $\pi:\A_\y\to\B(\H_\pi)$ be a *-representation of the
quasi-local algebra. It is convenient to prolong the system of
embedding maps into the representation, \ie we define
$i_\pila:\A_\la\to\B(\H_\pi)$ by $i_\pila=\pi\circ i_\yla$. For
uniformity of notation $\pi$ is sometimes also written as
$i_\piy$. There is a natural {\em ground state projection} in
$\H_\pi$, namely
$$ g_\pi= \slim_\la i_\pila(g_\la)
\quad,\deqno(gpi)$$
where the limit on the right hand side is in the strong operator
topology, and exists, because $i_\pila(g_\la)$ is a decreasing net
of projections by assumption \eq(g<g). Of course, the limit may be
zero.

Now let $\eta\in\B(\H_\pi)^*$ be a state with $\eta(g_\pi)=1$. Then
$\eta\circ\pi\in\SSG$. The converse, namely that
$\eta\circ\pi\in\SSG$ implies $\eta(g_\pi)=1$, is also true when
$\eta$ is normal, \ie continuous for the limit on the right hand
side of \eq(gpi). As a counterexample for singular $\eta$, consider
any faithful representation $\pi$ of $\A_\y$ in which every \bottom
state is singular or, equivalently, $g_\pi=0$. By Hahn-Banach
extension of $\omega$ from $\pi(\A_\y)$ to $\B(\H_\pi)$ we can write
$\omega=\eta\circ\pi$, with a necessarily singular state
$\eta\in\B(\H_\y)$. Then $\eta(g_\pi)=0$, because $g_\pi=0$, but
$\eta(i_\pila(g_\la))=\omega(i_\yla(g_\la))=1$, for all finite
$\la$.

We would like to prolong the inductive system of ground state
observables into the representation as well. Hence we define,
$$\eqalignno{
    r_\pi:\B(\H_\pi)&\to\B(\G_\pi)\cr
    r_\pi(A)&=g_\pi Ag_\pi
\quad,&\deqal(rpi)\cr
\noalign{\noindent
and, either for finite $\la$ or for $\la=\y$, }
    j_\pila:\B_\la&\to\B(\H_\pi)  \cr
    j_\pila(r_\la A)&:=r_\pi(i_\pila(A) )
\quad,\quad\hbox{for $A_\la\in\A_\la$.}
&\deqal(jpila)}$$
The salient facts about $j_\piy$ are collected in the following
Proposition.

\iproclaim/P.rep/Proposition.
\item{(1)}
The map $j_\piy$ is well-defined by equation \eq(jpila).
\item{(2)}
If $B_\la$ is $j$-convergent, then
$$ j_\piy(B_\y)=\slim_\la i_\pila(B_\la)
\quad,$$
where $\B_\la$ is considered as a subspace of $\A_\la$, and the limit is
in the strong operator topology.
\item{(3)}
If the inductive system has the product property, then $j_\piy$ is a
homomorphism.
\item{(4)}
When every $\omega\in\SSG$ is $\pi$-normal, $j_\piy$ is isometric.
\eproclaim

\proof:
(1) Suppose that $A_\la$ is $i$-convergent, with $r_\y(A_\y)=0$. That
is to say, \break
$\lim_\la\norm{g_\la A_\la g_\la}=0$. Then, since
$g_\pi\leq i_\pila(g_\la)$, and $i_\pila$ is a homomorphism, we have
$r_\pi\bigl(i_\pila(A_\la)\bigr)
  =g_\pi\bigl(i_\pila\bigl(g_\la A_\la g_\la\bigr)\bigr) g_\pi
  \to0$.

(2) It suffices to show this for basic nets of the form
$B_\la=g_\la i_\lalap(A_\lap)g_\la$, for some fixed
$A_\lap\in\A_\lap$. Then
$$ j_\piy(B_\y)=r_\pi(i_\piy i_\ylap (A_\lap))
              =g_\pi\ i_{\pi\lap}(A_\lap)\ g_\pi
\quad.$$
On the other hand,
$$i_\pila(B_\la)=i_\pila(g_\la)i_{\pi\lap}(A_\lap)i_\pila(g_\la)
\quad,$$
which converges strongly to the previous expression because, by
definition \eq(gpi), $g_\pi=\slim_\la i_\pila(g_\la)$, and because
the product is continuous for strong limits.

(3) Let $A_\la$, $B_\la$, and $C_\la=A_\la B_\la$ be $j$-convergent.
Then, by (2),
$$\eqalign{
  j_\piy(C_\y)&=\slim_\la i_\pila(A_\la B_\la)
               =\slim_\la i_\pila(A_\la)\  i_\pila(B_\la)         \cr
              &=\slim_\la i_\pila(A_\la)\ \slim_\la i_\pila(B_\la)
               =j_\piy(A_\y)\ j_\piy(B_\y)
\quad.}$$

(4) Let $A_\la$ be $i$-convergent, and $B_\la=r_\la A_\la$, and
recall that, up to norm small corrections, every $j$-convergent
$B_\la$ is of this form. Then
$$\norm{B_\y}=\sup_\Omega\abs{\Omega(B_\y)}
             =\sup_\Omega\abs{(r_\y^*\Omega)(A_\y)}
             =\sup_\omega\abs{\omega(A_\y)}
\quad,$$
where $\Omega$ runs over the unit sphere of $\B_\y^*$, and $\omega$
runs over all functionals in the linear hull of
$\SSG\subset\A_\y^*$ of norm $\leq1$.
Now by assumption all such functionals are represented as
$\omega(A_\y)=\tr(D_\omega \pi(A_\y))$, with $D_\omega$ a linear
combination of density matrices supported by $\G_\pi$, and of trace
norm $\leq1$. For such functionals, the last supremum is equal to
$\norm{g_\pi\pi(A_\y)g_\pi}=\norm{j_\piy(r_\y A_\y)}
      =\norm{j_\piy(B_\y)}$.
\QED

Condition (4) is by no means necessary to make $j_\piy$ isometric.
For example, if the product property holds, then it is sufficient
that $j_\piy$ is a faithful representation, whereas (4) requires
$j_\piy$ to be quasi-equivalent to the universal representation.
In fact, for our main example, we will construct a natural faithful
irreducible representation of $\B_\y$ arising from a representation
of $\A_\y$ (see Section 4.4). Whenever $j_\piy$ is faithful, we can
construct $\B_\y$ as $\B_\pi=g_\pi\,\pi(\A_\y)\,g_\pi$. This space is
then a C*-subalgebra of $\B(\H_\pi)$, but {\it not} of $\pi(\A_\y)$.

\bgsections  3. Two basic examples
\bgssection 3.1. The Heisenberg Ferromagnet

In this section we consider the \bottom states of a Heisenberg
ferromagnet on an arbitrary connected graph, with arbitrary positive
coupling constants, and with arbitrary (not necessarily equal) spins.
At each vertex $x$ of the graph we consider the Hilbert space
$\H_{\set x}=\Cx^{2s(x)+1}$ with an action of the spin-$s(x)$
representation $\DD{s(x)}$ of $\SU2$, with $s(x)>0$. The observable
algebra $\A_{\set x}$ is the algebra of operators on $\H_{\set x}$,
\ie the algebra of $(2s(x)+1)\times(2s(x)+1)$-matrices. The spin
operators in $\A_{\set x}$ will be denoted by $S_\alpha^x$,
$\alpha=1,2,3$. For larger regions $\la$ we set
$$ \H_\la=\bigotimes_{x\in\la}\H_{\set x}
\hskip50pt
   \A_\la=\bigotimes_{x\in\la}\A_{\set x}
\quad.\deqno(Hprods)$$
The injections $i_\lalap$ are given by
$i_\lalap(A)=A\otimes\idty_{\la\setminus\lap}$, as usual.
By abuse of notation we abbreviate the spin operators at vertex $x$,
considered as observables of the region $\la\ni x$ again as
$i_{\la,\set x}(S_\alpha^x)\equiv S_\alpha^x$.

For any finite, connected subset $\la$ of lattice vertices, the
Hamiltonian of the model is given by
$$ H_\la= -\sum_{\textstyle{{x,y\in\la}\atop x\frown y}}
          J_{xy} \sum_{\alpha=1}^3 S_\alpha^x S_\alpha^y
\quad,\deqno(HHeisen)$$
where ``$x\frown y$'' means that the vertices $x$ and $y$ are
connected by an edge, and $J_{xy}$ are arbitrary strictly positive
constants.

The operator $\sum_{\alpha=1}^3 S_\alpha^x S_\alpha^y$ commutes
with $\DD{s(x)}\otimes\DD{s(y)}$, and on the spin-$j$ subspace of
this representation it is equal to
$$ {1\over2}\bigl(j(j+1)-s(x)(s(x)+1)-s(y)(s(y)+1) \bigr)
\quad.\deqno()$$
This expression attains its maximum, namely $s(x)s(y)$, when
$j=s(x)+s(y)$ is the largest spin in the decomposition of
$\DD{s(x)}\otimes\DD{s(y)}$. Hence we have the lower bound
$$ \bra\phi\vert H_\la \vert \phi>
     \geq - \sum_{\textstyle{{x,y\in\la}\atop x\frown y}}
           J_{xy} s(x)s(y)
\quad,\deqno(HH>)$$
for any unit vector $\phi\in\H_\la$.
Clearly, this becomes an equality iff $\phi$ is supported by the
maximal spin subspace of $\H_{\set{x,y}}$ for any edge $x\frown y$.
Thus in order to compute the \bottom states of $H_\la$ we need to
analyze the intersection of maximal spin subspaces on overlapping
tensor factors.

\iproclaim/P.Heisen/Proposition.
Let $\la$ be a connected subgraph. Then the lowest eigenvalue of
$H_\la$ from \eq(HHeisen) is given by the right hand side of
\eq(HH>), and the corresponding eigenspace $\G_\la$ is the
irreducible subspace of $\bigotimes_{x\in\la}\DD{s(x)}$ for the
highest spin, $s_\la=\sum_{x\in\la}s(x)$.
\eproclaim

\proof:
Recall that the spin-$s$ representation $\DD s$ of $\SU2$ is
isomorphic to the subrepresentation of the $2s$-fold tensor power of
$\DD{1/2}$ on the completely symmetric subspace of
$(\Cx^2)^{\otimes 2s}$. Thus we can replace each site $x$ by a
collection $\widehat x$ of $2s(x)$ sites to each of which is
associated a Hilbert space $\Cx^2$ with a spin-$1/2$ representation
of $\SU2$. We are looking for the subspace of vectors $\Phi$ which are (a)
invariant the unitary operators exchanging any
two sites within the same cluster $\widehat x$, and (b) belong to
the highest spin subspace for the representation belonging to
$\widehat x\cup\widehat y$ for points $x,y$ connected by an edge.
Condition (b) simply means that the $\Phi$ is also symmetric for
with respect to the exchange of points from $\widehat x$ and
$\widehat y$. Since the graph is connected, such transpositions
generate the whole permutation group of the $2s_\la$ sites, and
$\Phi$ belongs to the completely symmetric subspace.
\QED

{}From this proof we can determine $\SSG$, using a Theorem of Hudson
and Moody \cite{HM}: it characterizes those states on an infinite
system, whose restriction to every finite subsystem is supported by
the completely symmetric (Bose) subspace, as the infinite products
of pure one-site states, and the integrals over such product states.
In our case, the one-site Hilbert space is $\Cx^2$, so the set of
pure states on a single site is naturally parametrized by a
$2$-sphere. Translated back to the language of spin systems with
arbitrary spin, we find that the extreme points of $\SSG$ are
characterized as those in which ``all the spins point in the same
direction''. To give a more precise description, let
$\chi_x\in\H_{\set x}$ denote the eigenvector of $S_3^x$ for the
largest eigenvalue $s(x)$. Then, for each $\la$, and each unit
vector $\uvec\in\Rl^3$, we set
$$\eqalign{
  \chi_\la&=\bigotimes_{x\in\la}\chi_x  \cr
  \chi_\la(\uvec)&=\bigotimes_{x\in\la} \DD{s(x)}_R \chi_x
\quad,}\deqno(Hprodvec)$$
where $R$ is a rotation taking the north pole into $\uvec$. Because
the sphere is the homogeneous space of $\SU2$ by the subgroup
generated by $S_3$, this vector does not depend on the choice of
$R$. Then the restriction of the extremal element of $\SSG$
belonging to $\uvec\in S$, restricted to a finite region $\la$ is
$B\mapsto\bra\chi_\la(\uvec),B_\la\chi_\la(\uvec)>$.

As a way of obtaining the \bottom states  this treatment is more
or less satisfactory. Some questions remain unclear, however. For
example, while we find that $\SSG$ is a simplex, \ie that every
$\omega\in\SSG$ has a unique integral decomposition into extreme
points, the nature of this simplex is less clear: is it a Bauer
simplex (with closed extreme boundary)  or a Poulsen simplex (with
dense extreme points), like the set of translationally invariant
states on a lattice (see Example 4.3.26 in \cite{BraRo})? The
difference between these two is precisely that the first kind of
simplex is the state space of a commutative C*-algebra, so we are
led to consider the space of observables with state space $\SSG$.
This is precisely the space $\B_\y$ which the inductive limit
construction of Section 2.2 yields naturally. Does it have a natural
algebraic structure in this case?

For computing the inductive limit, note that according to
\Prp/P.Heisen/, $\G_\la$ is the irreducible representation space of
$\SU2$ for spin $s(\la)=\sum_{x\in_\la}s(x)$, and all details of the
graph or the coupling constants become irrelevant. The embeddings
$j_\lalap$ are likewise independent of these details. An explicit
formula is the following: let
$$ V_\lalap:\Cx^{2s(\la)+1}\to \Cx^{2s(\lap)+1}\otimes
            \Cx^{2(s(\la)-s(\lap))+1}
$$
be the intertwining isometry between $\DD{s(\la)}$ and
$\DD{s(\lap)}\otimes\DD{s(\la)-s(\lap)}$, which is unique up to a
phase. Then
$$  j_\lalap(A)=V_\lalap^*(A\otimes\idty)V_\lalap
\quad.\deqno(Hjlalap)$$
Thus $j_\lalap$ depends only on the spins $s(\la)$ and $s(\lap)$, and
describes an inductive limit of the observable algebras on
irreducible representations of $\SU2$ with $s\to\infty$. Since the
half integer spin parameter is just the angular momentum in units of
$\hbar$, this limit is completely equivalent to the {\em classical
limit} $\hbar\to0$ for spins with fixed absolute value of angular
momentum \cite{Anja}.

{}From this perspective it would seem that the computation of $\B_\y$
can be based on asymptotic properties of Clebsch-Gordan
coefficients. However, it is more efficient to use the picture set
up in the proof of \Prp/P.Heisen/, and to exploit the high
permutation symmetry. This symmetry is at the root of the theory of
{\em Mean-field systems} \cite{MFRW,LMD,Delhi}.
This is, in fact, also the natural home for the Hudson-Moody
Theorem, as well as St\o rmer's more general non-commutative
analogue of the de Finetti Theorem \cite{Stormer}.
We briefly review the basic notions. Suppose that to each site $x$
in a finite set $\la$ we associate a Hilbert space $\H_{\set x}$ of
the same dimension. Then on $\H_\la=\bigotimes_{x\in\la}\H_{\set x}$
we have a natural action $\pi\mapsto U_\pi$ of the permutation group
of $\la$. The inductive limit underlying mean-field theory is given
by the algebras $\A_\la$ as above and the embedding maps
$$\eqalign{
      \sym_\lalap&:\A_\lap\to\A_\la   \cr
      \sym_\lalap(A)&={1\over\abs\la!} \sum_\pi
                U_\pi(A\otimes\idty_{\abs\la-\abs\lap})U_\pi^*
\quad,}\deqno(sym)$$
\ie a standard embedding $i_\lalap$, followed by an average over all
permutations over the larger set. It then turns out that the limit
space of the inductive system $(\A_\la,\sym_\lalap)$, which we
denote by $\A\MF$, is isomorphic to the space of continuous functions
on the state space $\SS(\A_{\set x})$ of the one-site algebra
\cite{MFRW}. One feature which carries over from this general
structure is the product property as defined in \Def/D.prod/:

\iproclaim/Hprod/Proposition.
The inductive limit defined by the maps \eq(Hjlalap),
has the product property. Moreover, the product induced on $\B_\y$
is commutative.
\eproclaim

\proof:
We realize $\G_\la$ as the completely symmetric subspace of
$\H_\la=\bigotimes_{x\in\la}\Cx^2$, \ie as the highest spin subspace
of the $\abs\la$-fold tensor product of the defining representation
of $\SU2$. Thus
$$ g_\la={1\over\abs\la!} \sum_\pi  U_\pi
\quad.$$
Because
$r_\la(U_\pi AU_\pi^*)
    =g_\la(U_\pi AU_\pi^*)g_\la
    =g_\la Ag_\la=r_\la(A)$,
an average over permutations is implicit in $r_\la$.
Therefore, we can write
$$ j_\lalap r_\lap(A)
    =r_\la i_\lalap(A)
    =r_\la\sym_\lalap(A)
\quad.$$
By a simple norm approximation argument it suffices to prove the
Proposition for basic nets $A_\la,B_\la$, \ie we can set
$$ A_\la= r_\la\sym_{\la\la_1}A_1
\qquad\hbox{and}\qquad
   B_\la= r_\la\sym_{\la\la_2}B_2
\quad,$$
for some fixed $A_1\in\A_{\la_1}$ and $B_2\in\A_{\la_2}$.
Since all symmetrized observables $\sym_\lalap X_\lap$ commute with
$g_\la$, we have
$$\eqalign{
   A_\la B_\la&=r_\la\Bigl((\sym_{\la\la_1}A_1)
                           (\sym_{\la\la_2}B_2) \Bigr)   \cr
       &=r_\la\Bigl(\sym_{\la,\la_1\sqcup\la_2}(A_1\otimes B_2)
              \Bigr) +\hbox{Rest}
\quad.\cr}$$
Here $\la_1\sqcup\la_2$ denotes the disjoint union of suitable
copies of $\la_1$ and $\la_2$. The first term in the last expression
is well-defined, since under symmetrization the labelling of sites
becomes irrelevant. The splitting of the last term is based on the
intuition that the product of symmetrized observables is an average
over products of copies of $A_1$ and $B_2$, permuted to localization
regions $\pi_1(\la_1)$ and $\pi_2(\la_2)$. As $\la$ becomes large,
these localization rarely intersect, and the term written is
(up to a small correction in normalization) the collection of terms
with $\pi_1(\la_1)\cap\pi_2(\la_2)=\emptyset$. The precise estimate
is given in Lemma IV.1 of \cite{MFRW}:
$$ \norm{\hbox{Rest}}\leq{\abs{\la_1}\cdot\abs{\la_2}\over\abs\la}
                     \norm{A_1}\cdot\norm{B_2}
     \buildrel\la\to\infty\over\longrightarrow  0
\quad.$$
Since the leading term is itself a basic net for the inductive
system $(\B_\la,j_\lalap)$, we have thus shown (1), and because this
term is the same for $B_\la A_\la$, we have shown (2).
\QED

As an abelian C*-algebra, $\B_\y$ is isomorphic to $\C(S)$ for
some compact space $S$. Its state space, which is topologically
isomorphic to $\SSG$ is therefore a Bauer simplex. It is also known
{}from the general mean-field theory that $S$ can be identified with a
set of homogeneous product states \cite{MFRW}, and it is easy to see
that only products of pure states have the Bose-Einstein symmetry,
which gives the Hudson-Moody Theorem \cite{Delhi}. We will not
describe these connections in detail, but instead use the above
identification \eq(Hprodvec) of the elements of $S$ to summarize the
result of this section.

\iproclaim/HBy/Theorem.
For the Heisenberg ferromagnet, we have $\B_\y\cong\C(S)$, where $S$
is the $2$-sphere. Under this isomorphism the limit $B_\y$ of a
$j$-convergent net $B_\la$ is the function defined by
$$ B_\y(\uvec)=\lim_\la \bra\chi_\la(\uvec),B_\la\chi_\la(\uvec)>
\quad.\deqno(Hjy)$$
\eproclaim

An interesting modification of this model is the chain with nearest
neighbour interaction
$$ H_2=-\sum_{ij}R_{ij}\sigma^i\otimes\sigma^j
\quad,\deqno(heirot)$$
where $R$ is a fixed $3\times3$ rotation matrix. Then by a
suitable $\SU2$-rotation at each site we can map the ground state
data of the interaction \eq(heirot) onto those of the Heisenberg
chain. The resulting inductive system is also isomorphic, and hence
the structure of the set of \bottom states is also the same. However,
as states on $\A_\y$ the states look quite different: if $R$ is a
rotation by an irrational angle, they will be almost periodic.

\bgssection 3.2. Valence bond solid states

We will now look at a construction that yields many examples of
locally exposed states, \ie \bottom state problems in which the face
$\SSG$ reduces to a single point. The typical example is a state
studied by Affleck, Kennedy, Lieb, and Tasaki \cite{AKLT}, namely
the unique ground state of the Hamiltonian
$$ H=\sum_x
   \left\{{1\over2}\vec S_x\cdot \vec S_{x+1}
       + {1\over 6}(\vec S_x\cdot \vec S_{x+1})^2
       +{1\over 3}\right\} \quad,
\deqno(aklt)$$
where $\vec S_x$ denotes the generators of the irreducible spin-1
representation of \SU2, acting in the one-site algebra at site $x$.
It is an anti-ferromagnetic model in the sense that it is an
increasing rather than a decreasing polynomial in the scalar product
of neighbouring spins. In fact, the expression in braces is nothing
but the projection onto the spin-2 subspace in the decomposition of
the tensor product of the two representations at sites $x$ and
$(x+1)$. Therefore, for any interval $\la\subset\Ir$, $\G_\la$ is
characterized by the property that on any two neighbouring sites the
total spin is  $\leq1$, whereas for the ferromagnetic ground state
only the maximal spin $2$ occurs.

In this special model the fastest way to determine the finite volume
ground states is the realization of the irreducible spin-$s$
representation as the space of homogeneous polynomials in two
variables of homogeneous degree $2s$. In this language it is easy to
see \cite{KLT,FCS} that the finite volume ground state spaces
$\G_\la$  are all four-dimensional, and contained in the spin$\leq1$
subspace of $\H_\la$. From an investigation of this model one can
abstract the following construction \cite{FCS,FCL}, which
no longer requires any symmetry group:

\iproclaim/D.VBS/ Definition.
A {\bf generalized valence bond solid {\rm(VBS)}} on a spin
chain with one-site Hilbert space $\H$ is given by
\item{(1)} two auxiliary finite dimensional
          Hilbert spaces $\K$ and $\Kbar$,
\item{(2)} a vector $\Phi\in\Kbar\otimes\K$
\item{(3)} a linear operator $S:\K\otimes\Kbar\to\H$.

\noindent
Then, for every interval $\la\subset\Ir$ of length $L$, the {\bf
ground state space} $\G_\la\subset\H_\la\equiv\H^{\otimes L}$ is
defined as the linear span of the set of vectors of the form
$$ \underbrace{S\otimes S\cdots\otimes S}_{L\ \rm factors}\
   \Bigl(\chi_L\otimes
    \underbrace{\phi\cdots\otimes\phi}_{L-1\ \rm factors}
      \otimes\chi_R \Bigr)
\quad,$$
with $\chi_R\in\K$ and $\chi_L\in\Kbar$ arbitrary.
A state $\omega$ whose restriction to each local subalgebra is
supported by $\G_\la$ is called a generalized {\bf VBS-state}.
\eproclaim

The point of this construction is that because
$\Phi\in\Kbar\otimes\K$ has an expansion into product vectors, the
condition \eq(g<g) is automatically satisfied, and, unless $S$
is somehow degenerate, \eq(g>0) also holds. Hence we have an
inductive limit of zero-energy observables in the sense of Section
2.2.

A fundamental observation in the theory of VBS states is that there
is a transfer matrix like operator, whose spectrum determines the
ground state degeneracy, and the decay properties of correlations in
the possible \bottom states. It leads to an alternative expression
for VBS states, which was introduced in \cite{FCS}, and studied in a
series of papers \cite{FCT,FCP,FCD,FQG} under the name of
{\em C*-finitely correlated states} (for an introduction, see also
\cite{FCL}. A copy of the construction was also made in \cite{MPG}).
The basic objects are the operators
$$\eqalign{
    V\ :\K  &\to \H\otimes\K  \cr
    V\chi   &= (S\otimes\id_\K)(\chi\otimes\phi)
\quad,\quad\hbox{and}\cr
   \E:\B(\K)&\to\B(\H)\otimes\B(\K)  \cr
   \E(X)    &=V^*XV
\quad. }\deqno(Vfcs)$$
For fixed $A\in\B(\H)$ we define a map $\E_A:\B(\K)\to\B(\K)$ by
$\E_A(B)=\E(A\otimes B)$. Then one easily verifies that, for each
interval of length $L$, the functionals on $\B(\H^{\otimes L})$
of the form
$$ \omega\bigl(A_1\otimes A_2\otimes\cdots\otimes A_L\bigr)
      =\rho\bigl(\E_{A_1}\E_{A_2}\cdots\E_{A_L}(B))
\quad,\deqno(fcs)$$
with $B\in\B(\K)$, and $\rho$ a linear functional on $\B(\K)$, span
the same space of functionals on $\B(\K)$ as those of the form
$A\mapsto\bra\psi_1,A\psi_2>$ with $\psi_1,\psi_2\in\G_L$. If $B$
and $\rho$ are positive, then the complete positivity of $\E$
implies that $\omega$ is also positive, and hence, with suitable
normalization is a state. {\em Correlation functions} in this state
are defined by setting $A_2=\cdots=A_{L-1}=\idty$ in \eq(fcs). Thus
on the right hand side we get powers of the linear operator
$\E_\idty:\B(\K)\to\B(\K)$. The spectral properties of this
{\em transfer operator} hence determine the behaviour of correlations.
As in the classical Frobenius theory of positive matrices,
$\E_\idty$ has a positive eigenvalue on its spectral radius
\cite{Albev}, and we say that $\E$ has {\em trivial peripheral
spectrum}, if this eigenvalue is simple, and all other eigenvalues
have strictly smaller modulus. By a simple transformation (see Lemma
2.5 in \cite{FCS}) one can take the Frobenius eigenvector to be the
identity element of the algebra, and $\E(\idty)=\idty$. Then, taking
$\rho$ in \eq(fcs) to be $\E$-invariant (\ie
$\rho\circ\E=\rho$), $\omega$ becomes normalized as a state for
every chain length $L$, and the states for different $L$ are the
restrictions of a unique translationally invariant state on the
quasi-local algebra. By construction, we have $\omega\in\SSG$.
If the eigenvalue $1$ of $\E_\idty$ is degenerate, then there are
several $\E$-invariant states $\rho$, and hence, in general, many
states in $\SSG$. Since $\E$ is an operator on the finite
dimensional space $\B(\K)$, the eigenspace of $1$ is finite
dimensional, so we expect $\SSG$ to be finite dimensional, too.
Proof of these intuitive statements can be found in $\cite{FCS}$.
Emphasis in that paper is on the case of trivial peripheral
spectrum. A detailed analysis of the degenerate case was undertaken
by Nachtergaele \cite{Bruno}. The proof of the following Theorem
draws on his results.

\iproclaim/T.fcs/Theorem.
Let $\G_\la, \la\subset\Ir$ be the ground state spaces of a valence
bond solid. Then
\item{(1)} $\SSG$ is a finite dimensional simplex, whose extreme
points are periodic pure states on $\A_\y$.
\item{(2)} If $\E$ has trivial peripheral spectrum, then $\SSG$
reduces to a single state.
\item{(3)} The inductive system $(\B_\la,j_\lalap)$ has the product
property.
\eproclaim

\proof:
(1) and (2) were proven in \cite{FCS}. Let $\omega^\alpha$,
$\alpha=1,\ldots,N$ denote the extreme points of $\SSG$, and let
$g_\la^\alpha\in\A_\la$ denote the support projection of the
restriction of $\omega^\alpha$ to $\A_\la$. It follows from
\cite{FCP} that the joint support projection of the $g_\la^\alpha$
coincides with $g_\la$ for large enough $\la$.
Then Nachtergaele
(\cite{Bruno}, Lemma 5) proves the estimate
$$ \lim_\la\norm{g_\la^\alpha\ i_\lalap(A)\ g_\la^\beta
              -\delta_{\alpha\beta}\ \omega^\alpha(A)\
                  g_\la^\alpha}
    =0
\quad.\deqno(Bruno)$$
Putting $A=\idty$ in this relation, we find that
$\widetilde g_\la=\sum_\alpha g_\la^\alpha$ is nearly a projection,
in the sense that
$\lim_\la\norm{(\widetilde g_\la)^2-\widetilde g_\la}=0$. Applying
the functional calculus, we find that
$\lim_\la\norm{g_\la-\widetilde g_\la}=0$. Hence, for fixed $\lap$,
and $A\in\A_\lap$,
$$ \lim_\la\norm{g_\la\ i_\lalap(A)\ g_\la-
       \sum_\alpha\omega^\alpha(A)g_\la^\alpha}=0
\quad.$$
Since $g_\la i_\lalap(A)g_\la=j_\lalap r_\lap(A)$ is a generic
$j$-convergent net, we find that the nets
$$ B_\la=\sum_\alpha\omega^\alpha(A)g_\la^\alpha
\deqno*()$$
also approximate every $j$-convergent net. Since the different
$\omega^\alpha$ are disjoint, varying $A$ yields arbitrary
coefficients $\omega^\alpha(A)$. Hence the $\la$-wise product of
nets of the form \eq*() is again of the same form, which proves (3).
\QED

\bgsections 4.   The infinite q-deformed Heisenberg ferromagnet
\bgssection 4.1. Definition of the model

In this subsection we derive the interaction \eq(Hq) as a quantum
group symmetric deformation of the Heisenberg chain. This is
helpful, for example, for seeing the ground state degeneracy ($L+1$
on the length $L$ chain) without computation. However, deformation
arguments are quite misleading with regard to the inductive limit.
Hence quantum groups will play no further role, and readers who are
not interested in this background can safely skip the rest of this
section.

The pair interaction of the spin-$1/2$ Heisenberg model considered
in Section 3.1 is the projection onto the spin-$0$ subspace in the
tensor product of two copies of the defining representation of
$\SU2$. The description of the model considered in this section is
exactly the same --- with $\SU2$ replaced by its quantum group
deformation $\SnU2$. We briefly recall some basic notions of quantum
group theory (following the approach of Woronowicz
\cite{WoRims,WoCMP}).

The structure of an ordinary (non-quantum) compact group $G$ can be
described completely in terms of the algebra $\C\equiv\C(G)$. The
topological structure of $G$ can be reconstructed from $\C$ vie the
Gelfand isomorphism, whereas the group structure can be encoded in
the coproduct $\copr:\C\to\C\otimes\C\cong\C(G\times G)$, given by
$(\copr F)(g_1,g_2)=F(g_1g_2)$. Associativity and existence of
neutral element and inverses can be reformulated in these terms as
well. The key observation leading to the theory of quantum groups is
that none of these axioms requires the commutativity of $\C$. Hence,
by definition (and modulo some important technical details
\cite{WoCMP}) a {\em quantum group} is a non-commutative C*-algebra
with coproduct $\copr$ satisfying all these axioms, except
commutativity of $\C$.

Basic notions of group theory are transferred to quantum groups by
the same principle. For example, a {\em matrix representation} of a
quantum group $(\C,\copr)$ is a matrix of finite dimension $d$, with
entries $u_{ij}\in\C$ such that
$$\copr(v_{ij})= \sum_{\ell=1}^d v_{i\ell}\otimes v_{\ell j}
\quad.\deqno()$$
The representation is called {\em unitary}, if this matrix is
unitary in the C*-algebra $\M_d(\C)\cong \M_d(\Cx)\otimes\C$ of
$d\times d$-matrices over $\C$. In particular, the defining, or
fundamental representation of $\SnU2$ is given by the matrix
$$  u=\left(\matrix{\alpha&-\qu\gamma^*\cr
                   \gamma&\alpha^*}\right)
\quad,\deqno()$$
where $\alpha$ and $\gamma$ are special elements in $\C$, and $\qu$,
with $0<\qu\leq1$ is the deformation parameter. This is the defining
representation also in the sense that the matrix elements $\alpha$
and $\gamma$ generate $\C$ as a C*-algebra. Unitarity of $u$ entails
the relations
$$\matrix{
      \alpha\alpha^*+\qu^2\gamma^*\gamma
   &=&\alpha\alpha^*+\qu^2\gamma\gamma^*
   &=&\alpha^*\alpha^*+\gamma\gamma^*
   &=&\idty \cr
 &&   \alpha\gamma^*-\qu\gamma^*\alpha
   &=&\alpha\gamma-\qu\gamma\alpha
   &=&0}
$$
The coproduct of $\SnU2$ is the *-homomorphism defined on
the generators by
$$\eqalign{
\copr\alpha &= \alpha\otimes\alpha- \qu\gamma^*\otimes\gamma \cr
\copr\gamma &= \gamma\otimes\alpha + \alpha^*\otimes\gamma
\quad.\cr}$$
Of course, for $\qu=1$ we obtain again the undeformed $\SU2$.

The {\em tensor product} of matrix representations is defined by
$$(v\otimes w)_{im,jn}= v_{ij}\ w_{mn}
\quad,\deqno(reptensor)$$
where the product on the right is the product in $\C$. It is
important to note that the non-commutativity of $\C$ introduces an
additional asymmetry here, \ie the tensor product defined as
$w_{mn}v_{ij}$ is really different from \eq(reptensor). A scalar
$k\time\ell$ matrix $V$ is called an {\em intertwiner} between the
matrix representations $v$ (of dimension $k$), and $w$ (of dimension
$\ell$), if
$$   \sum_n V_{in}v_{nm}=\sum_j w_{ij}V_{jm}
\quad.\deqno()$$
A {\em subgroup} of a quantum group is given by a quotient of $\C$
by a *-ideal, say $\ideal$, which is compatible with the coproduct in
the sense that
$\copr(\ideal)\subset\ideal\otimes\C+\C\otimes\ideal$. The quantum
group $\SnU2$ has a (non-quantum) subgroup corresponding to the
rotations around the $3$-axis. The corresponding *-ideal is generated
by the elements $\gamma$, and one readily verifies that the relation
$\gamma=0$ leaves the abelian algebra $\C/\ideal$ generated by a
unitary $\alpha$, with coproduct $\copr\alpha=\alpha\otimes\alpha$.

We use this observation to compute the $\SnU2$-invariant
interactions for the spin-$1/2$ chain, \ie the hermitian
intertwining operators $h$ between $(u\otimes u)$ and itself. This
is a straightforward calculation, which can be simplified
considerably by appeal to the known representation theory of $\SnU2$
\cite{WoRims}: the irreducible representations are labelled by a
half-integer spin parameter, and the decomposition of tensor
products yields the same irreducible blocks, \ie the same
Clebsch-Gordan series as $\SU2$. It follows that the space of
self-intertwiners of $(u\otimes u)$ is spanned by the identity and
one one-dimensional projection. This projection is the interaction we
are looking for; it projects onto the (up to a factor unique) vector
$\xi_\qu$ such that $(u\otimes u)\xi_\qu=\xi_\qu$. This is a
$\C$-valued equation, and by passing to the quotient defined by
$\gamma=0$ we immediately find that $\xi_\qu$ is of the form
$\xi_\qu=A\ket{+-}+B\ket{-+}$. Then we have, for example,
$$  \bra++\vert u\otimes u\vert\xi_\qu>
       = A\,u_{++}u_{+-}+B\,u_{+-}u_{++}
       = -\qu\lbrace\qu A+B\rbrace\ \gamma\alpha
       \buildrel!\over= \bra++\vert \xi_\qu>
       =0
\quad.$$
Hence $B=-\qu A$, which is confirmed in the other components of
$\xi_\qu$. To summarize, we consider the nearest neighbour
interaction $H^\qu_2$, where $H^\qu_2$ is the one-dimensional projection
onto the vector
$$    \xi_\qu= {1\over\sqrt{1+\qu^2}}
               \Bigl(\qu\ket{+-} -\ket{-+}\Bigr)
             \quad\in \Cx^2\otimes\Cx^2
\quad,\deqno(ximu)$$
where $\qu$ is a real parameter with $0\leq\qu<1$.
It is easy to check that the projection $H^\qu_2$ satisfies
Temperley-Lieb relations \cite{Levy,BatBar,Arndt}.

A direct characterization of such vectors without using quantum
groups is the following: the corresponding pure state has the
property that its marginals to the first and second factor (which
are mixed states) coincide. With a suitable choice of basis
(possibly alternating between odd an even sites) every vector of
this description can be written in the standard form \eq(ximu). More
general one-dimensional projections also define interactions with
many \bottom states. The analysis of such models can be carried out
along similar lines.

\bgssection 4.2. The finite chain

In this subsection we determine the \bottom states of the Hamiltonian
$$ H_\la=\sum_{x=1}^{L-1} i_{\la,\set{x,x+1}}(H_2^\qu)
\deqno()$$
on a chain of finite length $L$ with free boundary conditions.
The nearest neighbour interaction operator $H_2^\qu$ is as determined
in the previous section, namely the one-dimensional projection onto
the subspace generated by the ``$q$-deformed singlet''
$\qu\ket{+-}-\ket{-+}$. From the representation theory of $\SnU2$ it
is obvious that the ground state space will have the same dimension
as in the undeformed ($\qu=1$) case, namely $L+1$. However, we will
need more detailed information.

Let $\Psi(\sds1L)=\bra\sds1L\vert\Psi>$ denote the components of a
ground state vector. Then the condition
$$  \bigl(\idty_k\otimes H_2^\qu \otimes\idty_{L-k-2}\bigr)\Psi =0
\quad,$$
for some $0\leq k\leq (L-2)$ is equivalent to the condition that,
for arbitrary signs \break$\sds1k,\sds{k+3}L=\pm$,
$$ \Psi(\sds1k,-,+,\sds{k+3}L) = \qu \Psi(\sds1k,+,-,\sds{k+3}L)
\quad.\deqno()$$
Clearly, this determines each component $\Psi(\sds1L)$ in terms of
any other component with the same number $\np(\sds1L)$ of
``$+$''-signs. Hence the $(L+1)$ vectors
$$ \Phi_L(n)(\sds1L)=\delta_{n,\np(\sds1L)}\ \cdot\
                     \qu^{\sum_{x=1}^L\,
                      \textstyle x\ (1+\sigma_x)/2}
\deqno(phiL)$$
are an orthogonal, but not normalized basis of the ground state
space $\G_{\bracks{1,L}}$. The reason for choosing this particular
normalization is that, for $z\in\Cx$,
$$ \Psi_L(z)=\sum_{n=0}^L z^n\ \Phi_L(n)
\deqno(Psi:Phi)$$
becomes a product state: we have
$$\eqalign{
   \Psi_L(z)(\sds1L)&= \prod_{\textstyle{x=1\atop\sigma_x=+}}^L
                         z\qu^x\cr
   \Psi_L(z)&= \bigotimes_{x=1}^L \chi\bigl(z\qu^x\bigr)
\quad,}\deqno(Psiprod)$$
where, for any $z\in\Cx$, $\chi(z)\in\Cx^2$ denotes the vector with
components $\bra-\vert\chi(z)>=1$ and $\bra+\vert\chi(z)>=z$.
Since $\Psi_L(z)$ is the generating function for the $\Phi_L(n)$, it
is clear that these product vectors likewise span
$\G_{\bracks{1,L}}$. The norms of the vectors $\Phi_L(n)$ will be of
crucial importance. The following Lemma collects some basic formulas
and estimates.

\iproclaim/PhiLemma/ Lemma.
Define
$$\eqalign{
   \nphi(L,n)&=\qu^{-n(n+1)/2}\ \norm{\Phi_L(n)}
\quad,\quad\hbox{for $0\leq n\leq L<\infty$, and}    \cr
   \nphi(\infty,n)&=\lim_{L\to\infty}\nphi(L,n)
\quad.}$$
Then with $\ppp=\mfac\infty!$, and the convention $\mfac0!=1$, we
have, for $0\leq n\leq L\leq\infty$,
$$\eqalignno{\eqgroup()
          \nphi(L,n)&= {\mfac L! \over\mfac n! \, \mfac L-n!}
&\deqal\lasteq.a(nphi)\cr
\noalign{\vskip6pt}
     \nphi(L,n)     &=\nphi(L,L-n)
&\deqal\lasteq.b(nphisym)\cr
     \nphi(L,0)     &=1
&\deqal\lasteq.c()\cr
     \nphi(\infty,n)&=\prod_{i=1}^n(1-\qu^2)^{-1/2}
&\deqal\lasteq.d()\cr
     \ppp^{-1}&=\lim_{n\to\infty,\,L-n\to\infty}\nphi(L,n)
&\deqal\lasteq.e(nphiyy)\cr
           \ppp     \leq \nphi(L,n) &\leq \ppp^{-2}
\quad.&\deqal\lasteq.f(nphiest)
\cr}$$
\eproclaim

\proof:
Using the generating function \eq(Psi:Phi), and the orthogonality of
the $\Phi_L(n)$, we find, with the abbreviation $\lambda=\abs z^2$,
$$  \sum_{n=0}^L \lambda^n\,\nphi(L,n)^2\,\qu^{n(n+1)}
         = \prod_{i=1}^L(1+\lambda \qu^{2i})
\quad.\deqno*()$$
The expansion of this expression can be found in the literature on
$q$-factorials \cite{Gasp}, where $\nphi(L,n)^2$ appears as a $q$-deformed
binomial coefficient. This leads to \eq(nphi).
A direct verification uses \eq*() to obtain the recursion formula
$$ \nphi(L+1,n)^2=\nphi(L,n)^2+\qu^{2(L+1)} \nphi(L,n-1)^2
\quad,$$
which is satisfied by \eq(nphi).
The remaining properties are trivial consequences of \eq(nphi).
The proof of \eq(nphiest) uses the estimate.
$\ppp\leq \mfac n!\leq1$.
\QED

The importance of the estimate \eq(nphiest) is that it is uniform in
$L$, which is crucial for taking the limit to infinite chain
lengths. For taking norm estimates we will need to compute matrix
elements in an orthonormal basis, which is obtained by normalizing
the vectors $\Phi_L(n)$:
$$ \PHi_L(n)= \qu^{-n(n+1)/2}\,\nphi(L,n)^{-1}\, \Phi_L(n)
\quad.\deqno(PHi)$$

\bgssection 4.3. The Inductive Limit

The regions $\la$ indexing the net will be subintervals
of the integers. Each interval $\la\subset\Ir$ will be characterized
by two numbers $\la_\pm$ according to
$$ \la=\hint{\la_-,\la_+}
      =\Set{}{i\in\Ir}{\la_-<i\leq\la_+}
\quad.\deqno()$$
Thus $\la_+-\la_-$ is the number of sites in $\la$. We will use
$\la\to\y$ as shorthand for ``$\la_-\to-\infty$ and
$\la_+\to+\infty$''.

We will identify ground state spaces $\G_\la$ for intervals of the
same length, so we can write $\Phi_{\la_+-\la_-}(n)\in\G_\la$, with
the vectors defined in the previous subsection. Since we know that
$\G_L$ is spanned by product vectors, we can immediately write down
the isometry $V_\lalap$ identifying $\G_\la$ in the product of
ground state spaces for smaller chains:
$$\def\arrrray#1{\null\,\vcenter{\normalbaselines
    \ialign{\hfil$##\,$&$##$&\hfil$##$\hfil
           &&$\,\otimes\,$\hfil$##$\hfil\crcr
      \mathstrut\crcr\noalign{\kern-\baselineskip}
      #1\crcr\mathstrut\crcr\noalign{\kern-\baselineskip}}}\,}%
\arrrray{
    V_\lalap: \quad\G_\la\qquad
      &\longrightarrow
               &\G_{\hint{\la_-,\la'_-}}
               &\G_\lap
               &\G_{\hint{\la'_+,\la_+}}
\cr\noalign{\vskip4pt}
    V_\lalap \Psi_{\la_+-\la_-}(z)
      &=& \bigotimes_{i=\la_-+1}^{\la'_-} \chi(z\qu^i)
         &\bigotimes_{i=\la'_-+1}^{\la'_+} \chi(z\qu^i)
         &\bigotimes_{i=\la'_++1}^{\la_+} \chi(z\qu^i)
\cr\noalign{\vskip4pt}
      &=& \Psi_{\la'_--\la_-} (z)
         &\Psi_{\la'_+-\la'_-} \bigl(z\,\qu^{\la'_--\la_-}\bigr)
         &\Psi_{\la_+-\la'_+} \bigl(z\,\qu^{\la'_+-\la_-}\bigr)
\quad.}\deqno(V:psila)$$
Then it is straightforward to write down the maps $j_\lalap$
defined for general ground state spaces $\G_\la$ in Section 2:
$$j_\lalap(A)
     =V_\lalap^*\bigl(\idty_{\hint{\la_-,\la'_-}}\otimes
                       A     \otimes
                       \idty_{\hint{\la'_+,\la_+}}
                 \bigr)V_\lalap
\quad,\deqno(jla)$$
for $A\in\B_\lap\equiv\B(\G_\lap)$. Our aim is to study the
asymptotic behaviour of such expressions as $\la\to\infty$, up to
terms which become small in norm in this limit.

For developing norm estimates the product vectors in \eq(V:psila)
are not suitable. Therefore we begin by expressing $V_\lalap$ in the
orthonormal bases \eq(PHi). Inserting the generating function
\eq(Psi:Phi) into \eq(V:psila), collecting terms of the same
order in $z$ and expressing each $\Phi_L(n)$ in terms of its
normalized counterpart \eq(PHi), we find
$$\eqalignno{
    V_\lalap \PHi_{\la_+-\la_-}(n')
      &=\sum_{\ell' m' r'} \delta_{n',\ell'+m'+r'}\
        C_\lalap(\la'_- -\la_- -\ell',\ m'+\la'_-,\ r')
\quad\times\cr&\qquad\times\quad
        \PHi_{\la'_- -\la_-}  (\ell') \otimes
        \PHi_{\la'_+ -\la'_-} (m')\otimes
        \PHi_{\la_+  -\la'_+} (r')   \cr
      &=\sum_{\ell m r} \delta_{n-m,r-\ell}\
        C_\lalap(\ell,m,r)
        \quad \PHi_{\la'_- -\la_-}  (\la'_- -\la_- - \ell)
\quad\otimes\cr&\qquad\otimes\quad
        \PHi_{\la'_+ -\la'_-} (m-\la'_-)\otimes
        \PHi_{\la_+  -\la'_+} (r)
\quad,&\deqal(Vlalap)\cr\noalign{\noindent where}
  C_\lalap(\ell,m,r)
       &=\qu^{\ell(m-\la'_-) + \ell r +
              (\la'_+ -m)r}
\quad\times\cr\noalign{\vskip4pt}\cr&\qquad\times\quad
          {\nphi(\la'_- -\la_-  ,  \ell    )
           \nphi(\la'_+ -\la'_- , m -\la'_-)
           \nphi(\la_+  -\la'_+ , r        )
      \over\nphi(\la_+  -\la_-  , m+r-\ell-\la_-)}
\quad.&\deqal(Clalap)\cr}$$
The parameters $\ell',m',r'$ in the first sum in \eq(Vlalap) are the
numbers of ``$+$''-signs in the left, middle, and right segment of
the interval $\la$, respectively. However, these parameters are not
meaningful in the limit $\la\to\y$. Therefore the summation indices
in the second sum, and the arguments of $C_\lalap$ were chosen
slightly differently. They are the number $\ell=\la'_--\la_--\ell'$
of ``$-$''-spins on the right, the number $r=r'$ of ``$+$''-spins on
the right, and the label $m=m'+\la_-$ of the site, where ``$+$''
changes to ``$-$'' if we pack $m'$ ``$+$''-signs to the left of
$(\la'_+-\la'_--m')$ ``$-$''-signs into the interval $\la'$.
The ranges of these parameters are
$$ \matrix{ 0 &\leq& \ell &\leq& \la'_--\la_- \cr
      \la'_-  &\leq& m    &\leq& \la'_+       \cr
            0 &\leq& r    &\leq& \la_+-\la'_+ \cr}
\quad.\deqno()$$
As for the interval $\la'$ we will use for the whole interval $\la$
the parameter $n=n'+\la_-=(\ell'+m'+r')+\la_-=m+r-\ell$.

A convenient basis in $\B_\la$ is given by the usual matrix units,
parametrized as above. For any finite interval $\la$, and $\la_-\leq
n_1,n_2\leq\la_+$, we set
$$ E_\la(n_1,n_2)=\Ketbra\Big{\PHi_{\la_+-\la_-}(n_1-\la_-)}
                             {\PHi_{\la_+-\la_-}(n_2-\la_-)}
\quad.\deqno(Ela)$$
The operators play a dual role in the sequel. On the one hand,
because basic nets are dense, and the $E_\lap(m_1,m_2)$ are a basis
in $\B_\lap$, the limits of sequences of the form
$j_\lalap(E_\lap(m_1,m_2))$, with fixed $\lap, m_1,m_2$ span the limit
space $\B_\y$. On the other hand, they are convergent nets in their
own right (with $m_1,m_2$ fixed, and $\la\to\y$. In either case we
need the matrix elements of the operator
$j_\lalap(E_\lap(m_1,m_2))$, which at the same time  can be
considered as the matrix elements of the operator $j_\lalap$ itself.
Using \eq(Vlalap) we find
$$\eqalignno{
   J_\lalap&(n_1,m_1,m_2,n_2)
         \quad\equiv
    \Braket\Big{\PHi_{\la_+-\la_-}(n_1-\la_-)}
               {j_\lalap(E_\lap(m_1,m_2))}
               {\PHi_{\la_+-\la_-}(n_2-\la_-)}\cr
      &=\sum_{\textstyle{\ell_1r_1
                    \atop\ell_2r_2}}
           \delta_{n_1-m_1,\,r_1-\ell_1}\
           \delta_{n_2-m_2,\,r_2-\ell_2}\quad
           \delta_{\ell_1,\,\ell_2}\ \delta_{r_1,\,r_2}
\ \times\cr&\hskip20pt\times
   \overline{
        C_\lalap(\ell_1,\ m_1,\ r_1)} \quad
        C_\lalap(\ell_2,\ m_2,\ r_2)
\cr\noalign{\vskip10pt}
    &=\delta_{n_1-m_1,\, n_2-m_2}\ \sum_{\ell r}
      \delta_{n_1-m_1, r-\ell}\quad
        \overline{C_\lalap(\ell,\ m_1,\ r)} \quad
                  C_\lalap(\ell,\ m_2,\ r)
\quad.&\deqal(brajket)\cr}$$

In general, estimating operator norms is a difficult task. However,
in the present case we can utilize a special structure of the matrix
elements \eq(brajket): they are non-zero only along a single line
parallel to the main diagonal. The simple observation which allows
us to compute norms of such operators is stated in the following
Lemma.

\iproclaim/L.offdia/Lemma. Let $I\subset\Ir$ be a finite or infinite
subset, and $s\in\Ir$. Let $A$ be an operator in $\ell^2(I)$ whose
matrix elements $A_{n_1,n_2}$ with respect to the canonical basis
vanish unless $n_1-n_2=s$. Then
$$ \norm{A}=\sup_{n_1,n_2\in I}\abs{A_{n_1,n_2}}
\quad.$$
\eproclaim

\proof: The inequality ``$\geq$'' is trivial. For the converse, we
may take $I=\Ir$, by defining matrix elements $A_{n,m}=0$, if
$n\notin I$ or $m\notin I$. Thus $A=S\widetilde A$, with the
diagonal operator $\widetilde A$, defined by
$\widetilde A_{n,m}=A_{n+s,m}$, and a shift operator $S$.
Hence
$\norm{A}=\Norm{}{S\widetilde A}
         \leq\Norm{}{\widetilde A}
         =\sup\abs{\widetilde A_{n,m}}
         =\sup\abs{A_{n,m}}$.
\QED

\noindent
Our first key result is the convergence of the matrix units
themselves:

\iproclaim/L.Ela/Proposition.
Fix $m_1,m_2\in\Ir$. Then the net $E_\la(m_1,m_2)$
is $j$-convergent. Moreover, the net defined by
$$ \Fp_\la=\sum_{m=1}^{\la_+} E_\la(n,n)
\deqno(Fp)$$
is $j$-convergent.
\eproclaim

\proof:
(1) If we estimate each $\nphi(L,n)$ as in \eq(nphiest), a
straightforward estimate for $C_\lalap$ is
$$ \abs{C_\lalap(\ell,m,r)}
    \leq\ppp^{-7} \  \qu^{(\ell(m-\la'_-)+\ell r + (\la'_+-m)r)}
\quad.\deqno*()$$
For an upper bound on \eq(brajket) consider first the case
$n_1\geq m_1$. Then we have $r\geq1$ in the whole sum except,
possibly, in the term with $\ell=r=0$. Hence, apart from this term we
can estimate the above power of $\qu$ by
$\qu^{\ell  + (\la'_+-m_i))}$ ($i=1,2$) in every term with $\ell\geq1$.
The sum of these bounds is then a simple geometric series. Denoting
the term with $\ell=r=0$ by $J_\lalap^{00}(n_1,m_1,m_2,n_2) $, we
obtain the estimate
$$ \abs{J_\lalap(n_1,m_1,m_2,n_2)
      - J_\lalap^{00}(n_1,m_1,m_2,n_2) }
      \leq {\ppp^{-14}\over 1-\qu^2}\ \qu^{2\la'_+-m_1-m_2}
\quad.$$
An analogous estimate, with $(\la'_+-m_i)$ replaced by
$(m_i-\la'_-)$ applies for $n_1\leq m_1$. For
$J_\lalap^{00}(n_1,m_1,m_2,n_2)$ we get the explicit expression
$$ J_\lalap^{00}(n_1,m_1,m_2,n_2)
    = \delta_{n_1,m_1}\ \delta_{n_2,m_2}\
        {\nphi(\la'_+-\la'_-,m_1-\la'_-) \over
         \nphi(\la_+ -\la_- ,m_1-\la_-) }\
        {\nphi(\la'_+-\la'_-,m_2-\la'_-) \over
         \nphi(\la_+ -\la_- ,m_2-\la_-) }
\quad,\deqno(J00)$$
which converges to $\delta_{n_1,m_1}\delta_{n_2,m_2}$, as
$\la\to\y$, and then $\lap\to\y$. These are the matrix elements
of $\E_\la(m_1,m_2)$. Applying \Lem/L.offdia/, we get
$$\eqalign{
  \norm{j_\lalap(E_\lap(m_1,m_2))-E_\la(m_1,m_2)}
\hskip-140pt&\cr
      &\leq {\ppp^{-14}\over 1-\qu^2}\
           \max\Set\big:{\qu^{\la'_+-m_i},\ \qu^{m_i-\la'_\la}}
         + \sup_{n_1,n_2}
           \abs{\delta_{n_1,m_1}\delta_{n_2,m_2}
                  -J_\lalap^{00}(n_1,m_1,m_2,n_2) }  \cr
     &\longrightarrow0
\quad,\quad\hbox{as $\la\to\y$, and $\lap\to\y$.}
}$$

(2) By \Lem/L.offdia/ we have to show that, as $\la\to\y$, followed
by $\lap\to\y$,
$$\eqalign{
        \sum_{m\geq1} &J_\lalap(n,m,m,n)\longrightarrow 0
\quad,\quad\hbox{uniformly for $n\leq0$\quad, and }\cr
        \sum_{m\geq1} &J_\lalap(n,m,m,n)\longrightarrow 1
\quad,\quad\hbox{uniformly for $n\geq1$\quad.}
}$$
Since $j_\lalap\idty_\lap=\idty_\la$, the unrestricted sum over $m$
is equal to $1$ for all $n$, so the second statement is equivalent
to the convergence $\sum_{m\leq0} J_\lalap(n,m,m,n)\longrightarrow
0$, uniformly for $n\geq1$. By left/right symmetry proving this is
completely analogous to proving the first statement, so we will only
show the first.

Using again the estimate \eq*() in
\eq(brajket), we get
$$ \abs{ \sum_{m\geq1} J_\lalap(n,m,m,n)}
      \leq\ppp^{-14}\sum_{m\geq1}\sum_{\ell,\, r\geq0}
           \delta_{n-m,r-\ell}\
           \qu^{2\ell(m-\la'_-) + 2\ell r + 2(\la'_+ -m)r}
\quad.$$
Because $n\leq0$, the sum over $r$ begins at $r=0$, and
$\ell=r+(m-n)\geq(r+1)$. Hence, replacing the exponent of $\qu$ by
the smaller exponent $2r+2(m-\lap_-)$, we get the estimate
$$ \abs{ \sum_{m\geq1} J_\lalap(n,m,m,n)}
      \leq\ppp^{-14}\sum_{m\geq1}\sum_{ r\geq0}
           \qu^{2r+2(m-\lap_-)}
      =\ppp^{-14}(1-\qu^2)^{-2}\qu^{2(1-\la_-)}
\quad.$$
Clearly, this converges to zero as $\la_-\to-\infty$, uniformly in
$\la$ and $n$.
\QED

\iproclaim/unicon/Lemma. Fix $\lap$, and
$\lap_-\leq m_1,m_2\leq\lap_+$. Then
\item{(1)}
the limit
$$ J_\ylap(n_1,m_1,m_2,n_2)=\lim_\la  J_\lalap(n_1,m_1,m_2,n_2)
\deqno()$$
exists uniformly in $n_1$ and $n_2$.
\item{(2)}
The limits
$$\eqalign{
        \lim_{d\to+\infty} J_\lalap(d+m_1 ,m_1,m_2, d + m_2)
              &= \delta_{m_1,\,m_2}\delta_{m_1,\,\lap_-} \cr
        \lim_{d\to-\infty} J_\lalap(d+m_1 ,m_1,m_2, d + m_2)
              &= \delta_{m_1,\,m_2}\delta_{m_1,\,\lap_+} \cr
}\deqno(n->y)$$
exist uniformly in $\la$.
\eproclaim

\proof:
(1) In the proof of \Prp/L.Ela/ we have found a majorant for the sum
\eq(brajket), which is independent of $n_1$, $n_2$, and $\la$. Hence
it suffices to show that each term in \eq(brajket) goes to a limit
as $\la\to\y$, with fixed $n_1,n_2$. Hence the proof is completed
by the observation that, using \eq(Clalap) with \eq(nphiyy),
the limit
$$\eqalign{
    \lim_\la C_\lalap(\ell,m,r)
       &\equiv C_\ylap(\ell,m,r) \cr
       &= \qu^{\ell(m-\la'_-) + \ell r + (\la'_+ -m)r}\ \ppp
           \nphi(\y ,\ell)
           \nphi(\la'_+ -\la'_- , m -\la'_-)
           \nphi(\y, r )
\quad.}\deqno()$$
exists.

(2) As in the proof of \Prp/L.Ela/ we find, for $d\geq0$:
$$ \abs{J_\lalap(d+m_1,m_1,m_2,d+m_2)}
    \leq\ppp^{-14}(1-\qu^2)^{-1}\ \qu^{(2\lap_+-m_1-m_2)d}
\quad.$$
Hence, unless $m_1=m_2=\lap_+$ this goes to zero uniformly in $\la$.
In the exceptional case the only the first term ($r=\ell=0$) in the
sum \eq(brajket) survives, the remainder being bounded by a constant
depending only on $\qu$, times $\qu^{2d}$. That the first term
converges to $1$ follows again from \eq(J00). The statement for the
limit $d\to-\infty$ follows analogously.
\QED

\iproclaim/Edense/ Proposition.
For any  $j$-convergent net $B_\la\in\B_\la$, and $\epsilon>0$,
there is a finite linear combination
$$ B^\epsilon_\la=C\idty_\la + C_+\Fp_\la
               + \sum_{n_1,n_2}C_{n_1,n_2}E_\la(n_1,n_2)
\quad,$$
such that $\lim_\la\norm{A_\la-A^\epsilon_\la}\leq\epsilon$.
\eproclaim

\proof:
It suffices to consider basic nets of the form
$B_\la=j_\lalap(E_\lap(m_1,m_2))$ with fixed $\lap,m_1,m_2$. Assume
first that neither $m_1=m_2=\lap_-$ nor $m_1=m_2=\lap_+$. For any
$R\in\Nl$, let $P_\la^R$ denote the projection in $\B_\la$ onto the
subspace generated by the basis elements with $\abs n\leq R$. Then
by choosing $R$ sufficiently large, the norm difference
$\norm{B_\la - P_\la^R B_\la P_\la^R}$ can be made arbitrarily small,
uniformly in $\la$, on account of \Lem/unicon/(2), and
\Lem/L.offdia/.  We now introduce
$$ B_\la^R=\sum_{\abs{n_i}\leq R} J_\ylap(n_1,m_1,m_2,n_1)
            E_\la(n_1,n_2)
\quad.$$
Then $\norm{P_\la^R B_\la P_\la^R-B_\la^R}$ converges to zero as
$\la\to\y$, because of the first part of \Lem/unicon/. Moreover,
$B_\la^R$ is a finite linear combination of nets of the form
\eq(Ela), and therefore satisfies the conditions of the Proposition
for sufficiently large $R$. In the two exceptional cases, the same
arguments apply after subtraction of either $\Fp_\la$ or
$(\idty-\Fp_\la)$ from both nets.
\QED

\iproclaim/Limit/Theorem.
The inductive system \eq(jla) has the product property.
$\B_\y$ is the C*-algebra of operators on $\ell^2(\Ir)$, generated
by the compact operators, the identity, and the operator of
multiplication with the characteristic function of $\Nl\subset\Ir$.
\eproclaim

\proof:
It suffices to show that the nets of the form $\B^\epsilon_\la$ as
in \Prp/Edense/ have the product property. However, this is evident
{}from the multiplication rule
$$ E_\la(m_1,m_2)E_\la(m_3,m_4)
        =\delta_{m_2,m_3}E_\la(m_1,m_4)
$$
for matrix units, which holds for every $\la$.
Hence $\B_\y$ is the unique C*-algebra generated by elements
$E_\y(m_1,m_2)$, $\Fp_\y$, and $\idty$ with just these multiplication
rules.
\QED

\noindent
Combining this with \Thm/iso/ we now obtain the convex structure of
$\SSG$.

\iproclaim/Statesp/Corollary.
$\SSG$ is isomorphic to the convex hull of the set of density
matrices on $\ell^2(\Ir)$, and two additional points $\omega_+$ and
$\omega_-$, interpreted as the ``all spins up'' and the ``all spins
down state''.
\eproclaim

Note that \Thm/Limit/ gives additional information about the
topology of $\SSG$, which is not contained in the above description
of the convex structure.
In fact, there is a different C*-algebra with the same convex set as
its state space. Since any C*-algebra $\C$ is abstractly
reconstructed as the set of $\sigma(\C^*,\C)$-{\it continuous}
affine functionals on its state space $\SS(\C)$, this shows that the
description of $\SSG$ as a mere convex set without topology misses a
vital element. The second C*-algebra with the same convex state
space is the (C*-algebraic) direct sum of a one-dimensional algebra
and the algebra of compact operators with identity adjoined. The
projection onto the first summand then produces a continuous affine
functional which is $1$ on the state, say, $\omega_*$, and vanishes
on the non-abelian part. In other words, $\omega_*$ is an exposed
state. In contrast to this, neither $\omega_-$ nor $\omega_+$ is
exposed in $\SS(\B_\y)\cong\SSG$.

The program for computing $\SSG$ can be carried out slightly more
easily for the same model on the half chain \cite{Claudius}. In that
case analogous results hold, but only one state at infinity needs to
be adjoined.

\bgssection 4.4. Representation in an infinite tensor product

For the determination of the inductive limit in the previous section
we did not use any particular representation of the quasi-local
algebra. Indeed, we argued in Section 2.3 that this is essential for
obtaining a complete characterization of $\SSG$, and not just the
subset of states which happen to be normal in the given
representation. On the other hand, quantities like the limiting
matrix elements $J_\ylap(n_1,m_1,m_2,n_2)$ appearing in the
calculation have a natural interpretation in a Hilbert space
associated with the infinite system. In this section we make this
connection explicit.

The starting point is that the product vectors $\Psi_L(z)$, which
span the ground state space for every chain length $L$, contain
mostly ``spin up'' factors on the left, and ``spin down'' factors on
the right. Therefore, infinite product vectors, which should represent
ground states of the infinite chain, should make sense in the
incomplete tensor product \cite{Gui}
$$ \H_\pi=\bigotimes_{i\in\Ir}(\Cx^2,\eta_i)
\quad,\deqno(Hpi)$$
with the reference vectors
$$ \eta_i=\cases{ {0\choose1} & for $i>0$, and
                 \cr\noalign{\vskip4pt}
                  {1\choose0} & for $i\leq0$. \cr}
\deqno(refvec)$$
This space carries an irreducible representation
$\pi:\A_\y\to\B(\H_\pi)$ of the quasi-local algebra, such that
$\pi(A_i)$ for $A_i\in\A_{\set i}$ acts in the $i\th$ tensor factor.
Of course, since $\A_\y$ is simple, this representation is faithful.

The two types of vectors we used above for every finite $\la$,
namely the product vectors $\Psi$, and the orthogonal vectors
$\Phi$, can be embedded into $\H_\pi$ as follows.
For $z\in\Cx$, we define
$$ \chi_i(z)= \cases{ {z\choose 1} &$=\chi(z)$ for $i>0$, and
                     \cr\noalign{\vskip4pt}
                      {1\choose1/z}&$=z^{-1}\chi(z)$ for $i\leq0$,
                      \cr}
\deqno(chiz)$$
where $\chi(z)$ is defined as in \eq(Psiprod). For finite $\la$
we now define the product vectors
$$\eqalignno{
  \Psii_\la(z)&=\bigotimes_{i\in\la}\chi_i(z\qu^i)
                \otimes\bigotimes_{i\notin\la}\eta_i
\quad,&\deqal(Psii)\cr
\noalign{\noindent and the orthogonal vectors}
 \Phii_\la(n)&=\Phi_{\la_+-\la_-}(n-\la_-)
                \otimes\bigotimes_{i\notin\la}\eta_i
\quad,&\deqal(Phii)
}$$
where $\la_-<n\leq\la_+$. The basic property of these vectors is
stated in the following Theorem:

\iproclaim/T.ipro/Theorem.
The limits $\Phii_\y(n)=\lim_\la\Phii_\la(n)$ and
$\Psii_\y(z)=\lim_\la\Psii_\la(z)$ exist in the norm of $\H_\pi$ for
all $n\in\Ir$, and all $z\in\Cx\setminus\set0$. The $\Phii_\y(n)$ are
an orthonormal basis of $\G_\pi$, and, for $z\in\Cx\setminus\set0$,
we have the convergent expansion
$$ \Psii_\y(z)=\ppp^{-1} \sum_{n=-\infty}^\infty z^n\
                \qu^{n(n+1)/2}\ \Phii_\y(n)
\quad.\deqno()$$
A \bottom state $\omega\in\SSG$ is normal in the representation $\pi$
if and only if it is disjoint from both $\omega_+$ and $\omega_-$,
in which case it is given by a unique density matrix supported by
$\G_\pi$.
\eproclaim

\proof:
The convergence of $\Psii_\la(z)$ follows from the standard theory of
incomplete tensor products \cite{Gui}, because
$$ \sum_{i\in\Ir} \norm{\eta_i-\chi_i(z\qu^i)}^2\leq\infty
\quad.\deqno(Psiiconv)$$
Since every $\Phii_\la(n)$ is a unit vector,
$$\norm{\Phii_\la(n)-\Phii_\lap(n)}^2
    =2-2\Re\bra\Phii_\la(n),\Phii_\lap(n)>
\quad.$$
The scalar product is computed in
$\G_{\hint{\la_-,\la'_-}}\otimes\G_\lap
          \otimes\G_{\hint{\la'_+,\la_+}}$
into which $\Phii_\la(n)$ is embedded as in \eq(Vlalap), and
$\Phii_\lap(n)$ by tensoring with suitable spin up, resp.\ spin down
vectors:
$$\eqalign{
  \bra\Phii_\la(n),\Phii_\lap(n)>
\hskip-50pt
&\cr
     &=\bra V_\lalap\PHi_{\la_+ -\la_-}(n-\la_-),
                    \PHi_{\la'_- -\la_-}(\la'_- -\la_-)\otimes
                    \PHi_{\la'_+ -\la'_-}(n-\la'_-)  \otimes
                    \PHi_{\la_+ -\la'_+}(0)  >  \cr
     &=C_\lalap(0,n,0)
      ={\nphi(\la'_+ -\la'_- , n-\la'_-)  \over
        \nphi(\la_+ -\la_- , n-\la_-)  }
\quad.}$$
Since numerator and denominator both converge to $\ppp$, we have
$\lim_\lap\lim_\la C_\lalap(0,n,0)=1$, and $\Phii_\la(n)$ is a Cauchy
net in $\H_\pi$.

For different $n$ the vectors $\Phii_\y(n)$ are clearly orthogonal.
They span $\G_\pi$, because every vector in $\H_\pi$ is approximated
by vectors differing from the reference vector only in a finite
region sites $\la$, and because for approximating a given vector
$\Phii\in\G_\pi$ we may apply the projection
$i_\pila(g_\la)\leq g_\pi$ to the approximating vectors without
loss.

The expansion formula is obtained from the corresponding formula for
finite $\la$. Assuming $\la_-\leq0$ and using, in succession,
equations \eq(chiz), \eq(Psiprod), \eq(Psi:Phi), and \eq(PHi),  we get
$$\eqalign{
    \bigotimes_{i\in\la} \chi_i(z\qu^i)
        &= \prod_{i=\la_-+1}^{0}(z\qu^i)^{-1}\
                  \bigotimes_{i\in\la} \chi(z\qu^i)    \cr
        &= z^{\la_-}\ \qu^{(\abs{\la_-}-1)\abs{\la_-}/2}\
               \Psi_{\la_+ -\la_-}(z\qu^{\la_-})        \cr
        &= z^{\la_-}\ \qu^{(\la_- +1)\la_- /2}\
           \sum_{n'=0}^{\la_+ -\la_-} (z\qu^{\la_-})^{n'}\
               \Phi_{\la_+ -\la_-}(n')        \cr
        &= \sum_{n=\la_-}^{\la_+} z^{n}\
               \nphi(\la_+ -\la_-, n -\la_-)\ \qu^{n(n+1)/2}\
               \PHi_{\la_+ -\la_-}(n-\la_-)
\quad.\cr}$$
Multiplying with the appropriate tensor product of reference vectors
$\eta_i$, we get
$$ \Psii_\la(z)= \sum_{n=\la_-}^{\la_+} z^{n}\
               \nphi(\la_+ -\la_-, n -\la_-)\ \qu^{n(n+1)/2}\
               \Phii_{\la}(n)
\quad,$$
and the expansion for infinite $\la$ follows, because
$\nphi(\la_+ -\la_-, n -\la_-)\longrightarrow1$.
Note that the sum converges both for $n\to+\infty$ and
$n\to-\infty$, and arbitrary $z\neq0$, because of the quadratic
dependence of the exponent of $\qu$ on $n$.
\QED

The two $\pi$-singular states $\omega_\pm$ are clearly the limits of
any  $\pi$-normal state, for shifts to $\pm\infty$. Hence {\em all}
states are of the form $\eta\circ\pi$ for a (possibly singular)
state $\eta\in\SS(\B(\H))$. Stating this observation  as a property
of $j_\piy$, and using the product property, we get

\iproclaim/P.cluster/Proposition.
$j_\piy:\B_\y\to\B(\G_\pi)$ is a faithful irreducible representation.
\eproclaim

\bgssection 4.5. The spectral gap

In the representation $\pi$ we define the Hamiltonian $H_\pi$ as the
closure of the operator given by
$$  H_\pi\ i_{\pi\lap}(A) \phi
        = \lim_\la\ \bracks{ \pi(H_\la),i_{\pi\lap}(A)}\  \phi
\quad,\quad\hbox{for $A\in\A_\lap$, and $\phi\in\G_\pi$.}$$
Here the net on the right hand side is eventually constant, because
$A\in\A_\lap$ is strictly local, and the interaction is finite range.
It is a standard argument \cite{BraRo} to show that the dynamics
exist as an automorphism on the quasi-local algebra and is generated
by $H_\pi$. Hence $H_\pi$ has a dense set of analytic vectors, and
is essentially self-adjoint. For a pure VBS state it was shown that
$H_\pi$ has a spectral gap above its unique ground state. These
results were extended to general VBS states by \cite{Bruno}. On the
other hand, for the Heisenberg ferromagnet the well-known magnon (or
spin wave) states have arbitrarily small energy, and hence $H_\pi$
has no gap. Here we will show that even an arbitrarily small
deformation generates a gap. The basic technique for obtaining lower
estimates on a gap is given in the following Lemma, which was proved
in \cite{FCS} (see the proof of Theorem 6.4 in that paper). Some
refinements and generalizations are in \cite{Bruno,Brunogap}.

\iproclaim/L.gap/Lemma.
Consider a one-dimensional translationally invariant nearest
neighbour interaction, whose ground state projections $g_\la$
satisfy assumption \eq(g<g). For $L\in\Nl$ let $\gamma_L$ denote the
gap of
$H_{\hint{0,L}}$, \ie the largest number satisfying
$$ H_{\hint{0,L}}\geq \gamma_L (\idty-g_{\hint{0,L}})
\quad.\deqno()$$
For $p\in\Nl$, consider the numbers
$$ \epsilon(p)=\norm{(g_{\hint{0,2p}}\otimes\idty^{\otimes p})
                     (\idty^{\otimes p}\otimes g_{\hint{p,3p}})
                     -g_{\hint{0,3p}} }
\quad.\deqno()$$
Then, for $n\geq3$:
$$ \gamma_{n\cdot L} \geq {\gamma_{2L}\over2}(1-2\,\epsilon(p))
\quad.\deqno(gapest)$$
\eproclaim

The quantities appearing in this Lemma are readily computable for
small chains. With some assistance for doing long symbolic
computations \cite{math} we find for our model:
$$\eqalign{
    \gamma_2&=1  \cr
    \gamma_3&=(1 - \qu + \qu^2)/(1 + \qu^2)   \cr
    \gamma_4&=(1 - \sqrt2\, \qu + \qu^2)/(1 + \qu^2)\cr
     \epsilon_1&=\qu/(1 + \qu^2)  \cr
     \epsilon_2&=\qu^2/(1 + \qu^4)
\quad.\cr}$$
This suffices to determine the first two bounds resulting from
\Lem/L.gap/:
$$\eqalignno{
  \hbox{bound}(p=1)&={(1-\qu)^2 \over 2(1 + \qu^2)} \cr
  \hbox{bound}(p=2)&={(1-\qu^2)^2\ (1 - \sqrt2\qu + \qu^2)\over
                      2\,(1 + \qu^2)\ (1 + \qu^4)}
}$$
Here the second bound turns out to be only a slight improvement over
the first. In any case, both bounds go to zero as $\qu\to1$, which
is also clear from the existence of low lying magnon excitations in
the undeformed model.

\let\REF\doref
\Acknow
We would like to thank Aernout van Enter for stimulating
discussions, and many hints about relevant literature. We would also
like to thank Bruno Nachtergaele for many helpful comments, and for
making his work \cite{Bruno,Brunogap} available prior to publication.

\REF AKLT AKLT      \Jref
     I. Affleck, T. Kennedy, E.H. Lieb, H. Tasaki
     "Valence bond ground states in isotropic quantum
     antiferromagnets"
     Commun.Math.Phys. @115(1988) 477--528

\REF AKS Korepin \Gref
    G. Albertini, V.E. Korepin, A. Schadschneider
    "Long string as ground state of the XXZ model with fixed
    magnetization"
    Preprint ITP-SB-94-45, Stony Brook 1994,
    archived as {\tt cond-mat/9411051}

\REF AHK Albev \Jref
    S. Albeverio, R. H\o egh-Krohn
    "Frobenius theory of positive maps of von Neumann algebras"
    Commun.Math.Phys. @64(1978) 83--94

\REF ASW  Wres \Gref
    F.C. Alcaraz, S.R. Salinas, W.F. Wreszinski
    "Anisotropic ferromagnetic quantum domains"
    Preprint, Sao Paulo 1994

\REF AB$^{\bf 3}$Q AlBaBaBaQ \Jref
    F.C. Alcaraz, M.N. Barber, M.T. Batchelor, R.J. Baxter, G.R.W.
    Quispel
    "Surface exponents of the quantum XXZ, Ashkin--Teller and Potts
    models"
    J.Phys. @A20(1987)6397--6409

\REF Alf Alfsen \Bref
    E.M. Alfsen
    "Compact convex sets and boundary integrals"
    Springer-Verlag, Berlin 1971

\REF AHY Arndt \Gref
    P.F. Arndt, T. Heinzel, C.M. Yung
    "Temperley-Lieb words as valence bond ground states"
    Preprint Bonn HE-94-26, archived as  {\tt cond-mat/9411085}

\REF BB BatBar \Jref
    M.T. Batchelor, M.N. Barber
    "Spin-s quantum chains and Temperley-Lieb algebras"
    J.Phys. @A23(1990) L15--L21

\REF BY BY \Gref
     M.T. Batchelor, C.M. Yung
     "q-Deformations of quantum spin chains with exact valence bond
      ground states"
      preprint archived as {\tt cond-mat/9403080}

\REF Bax Baxter \Bref
    R.J. Baxter
    "Exactly solved models in statistical mechanics"
    Academic Press, London 1982

\REF BR BraRo      \Bref
    O. Bratteli, D.W. Robinson
     "Operator algebras and quantum statistical mechanics"
     2 volumes, Springer Verlag, Berlin, Heidelberg, New York
     1979 and 1981

\REF Dri Drinfeld \Gref
    V.G. Drinfel'd
    "Quantum groups"
    pp. 798-820 in vol.1 of the Proceedings of the Int.Congr.Math.
    Berkeley 1986,
    Academic Press 1987

\REF DW1 MFD   \Jref
    N.G. Duffield,  R.F. Werner
    "Mean field dynamical semigroups on C*-algebras"
    Rev.Math.Phys. @4(1992) 383--424

\REF DW2 LMD   \Jref
    \sameauthor. {}
    "Local dynamics of mean-field quantum systems"
    Helv.Phys.Acta @65(1992) 1016--1054

\REF EFS Ave \Jref
    A.C.D. van Enter, R. Fern\'andez, A.D. Sokal
    "Regularity properties and pathologies of position-space
    renormalization-group transformations:
    scope and limitations of Gibbsian theory"
    J.Stat.Phys. @72(1993)879--1167

\REF FNW1 FCS      \Jref
    M. Fannes, B. Nachtergaele, R.F. Werner
    "Finitely correlated states of quantum spin chains"
    Commun.Math.Phys. @144(1992) 443--490

\REF FNW2 FCT \Jref
    \sameauthor. {}
    "Ground states of VBS models on Cayley trees"
    J.Stat.Phys. @66(1992) 939--973

\REF FNW3 FCP      \Jref
    \sameauthor. {}
    "Finitely correlated pure states"
    J.Funct.Anal. @120(1994) 511--534

\REF FNW4 FCD \Jref
    \sameauthor. {}
    "Abundance of translation invariant pure states
    on quantum spin chains"
    Lett.Math.Phys. @25(1992) 249--258

\REF FNW5 FQG \Gref
    \sameauthor. {}
    "Quantum spin chains with quantum group symmetry"
    Preprint Leuven, KUL-TF-94/8

\REF FYF Frahm \Jref
    H. Frahm, N.-C. Yu, M. Fowler
    "The integrable XXZ-Heisenberg model with arbitrary spin:
    construction of the Hamiltonian, the ground-state configuration
    and conformal properties"
    Nucl.Phys. @B336(1990)396--434

\REF FP FroPfi \Jref
    J. Fr\"ohlich, C.-E. Pfister
    "Spin waves, vortices, and the structure of equilibrium states
    of the classical XY model"
    Commun.Math.Phys. @89(1983)303--327

\REF GR   Gasp  \Bref
    G. Gasper, M. Rahman
    "Basic hypergeometric series"
    Encyclopedia of Mathematics and its applications (Vol.35),
    Cambridge University Press, Cambridge 1990

\REF G\'om  Gomez \Jref
    G. G\'omez-Santos
    "Role of domain walls in the ground-state properties of the spin
    -1/2 XXZ Hamiltonian on the linear chain"
    Phys.Rev. @B41(1990)6788--6793

\REF Got Claudius \Gref
    C.-T. Gottstein
    "Lokal exponierte Zust\"ande auf Quantenspinketten"
    Diplom Thesis, Osnabrck Sept.\ 1994

\REF Gui   Gui  \Bref
    A. Guichardet
    "Tensor products of $C^*$-algebras"
     Lecture Notes Series (No.12 + 13),  Aarhus Universitet
     Aarhus 1969

\REF HMRS Scheunert \Jref
    H. Hinrichsen, P.P. Martin, V. Rittenberg, M. Scheunert
    "On the two-point correlation functions for the U$_q$[SU(2)]
     invariant spin one-half Heisenberg chain at roots of unity"
    Nucl.Phys. @B415[FS](1994)533--556

\REF HS HSlawny \Jref
    W. Holsztynski, J. Slawny
    "Peierls condition and number of ground states"
    Commun.Math.Phys. @61(1978) 177--190

\REF  HM  HM  \Jref
    R.L. Hudson,  G.R. Moody
    "Locally normal symmetric states and an analogue of de Finetti's
    Theorem"
    Z.Wahrsch.theorie.verw.Gebiete @33(1976) 343--351

\REF JMMN Jimbo \Jref
    M. Jimbo, K. Miki, T. Miwa, A. Nakayashiki
    "Correlation functions of the XXZ model for $\Delta<-1$"
    Phys.Lett.A @168(1992)256--263

\REF Joh Johnson \Jref
    J.D. Johnson
    "A survey of analytic results for the 1D Heisenberg magnets"
    J.Appl.Phys. @52(1981)1991--1992

\REF KLT KLT \Jref
    T. Kennedy, E.H. Lieb, H. Tasaki
    "A two-dimensional isotropic quantum antiferromagnet with unique
    disordered ground state"
    J.Stat.Phys. @53(1988) 383--415

\REF  KSZ MPG \Jref
    A. Kl\"umper, A. Schadschneider, J. Zittartz
    "Matrix product ground states for one-dimensional spin-1
    quantum antiferromagnets"
    Europhys.Lett. @24(1993)293--297

\REF KT KTasaki \Jref
    T. Koma, H. Tasaki
    "Symmetry breaking and finite-size effects in quantum many-body
    systems"
    J.Stat.Phys. @76(1994)745--803

\REF Lev Levy \Jref
    D. Levy
    "Algebraic structure of translation-invariant spin-1/2 xxz and
    q-Potts quantum chains"
    Phys. Rev. Lett. @67(1991)1971--1974

\REF Mat math \Bref
   \noinitial. Wolfram Research{,} Inc.
   "Mathematica 2.2"
    Wolfram Research, Inc., Champaign, Illinois 1992

\REF Men Anja \Gref
    A. Menkhaus
    "Der klassische Limes fr Spinsysteme"
    Diplom Thesis Osnabrck, in preparation

\REF MN MezNep \Jref
    L. Mezincescu, R.I. Nepomechie
    "Analytical Bethe Ansatz for quantum-algebra-invariant spin
    chains"
    Nucl.Phys. @B372(1992) 597--621

\REF Mie Miek \Jref
    J. Mi\c ekisz
    "The global minimum of energy is not always a sum of local
    minima --- a note on frustration"
    J.Stat.Phys. @71(1993) 425--434

\REF Na1 Bruno \Gref
    B. Nachtergaele
    "The spectral gap for some spin chains with discrete symmetry
    breaking"
    Preprint, Princeton 1994
    archived as {\tt cond-mat/9410110}

\REF Na2 Brunogap \Gref
    B. Nachtergaele
    "A lower bound for the spectral gap of the ferromagnetic XXZ
    chain"
    Preprint Princeton, archived as {\tt condmat/9501098}

\REF Nag NAG  \Gref
    R.J. Nagel
    "Order unit and base norm spaces"
    \inPr  A. Hartk\"amper,  H. Neumann
    "Foundations of quantum mechanics and ordered linear spaces"
    Lecture Notes in Physics 29,
    Springer Verlag, Berlin, Heidelberg, New York 1974

\REF PS Pasq \Jref
    V. Pasquier, H. Saleur
    "Common structures between finite systems and conformal field
     theories through quantum groups"
    Nucl.Phys.B @330(1990)523--556

\REF Pec Pechersky \Jref
    E.A. Pechersky
    "The Peierls condition (or GPS condition) is not always
    satisfied"
    Sel.Math.Sov. @3(1983/84)87-91

\REF   RW  MFRW        \Jref
    G.A. Raggio,  R.F. Werner
    "Quantum statistical mechanics of general mean field systems"
    Helvet.Phys.Acta.  @62(1989)980--1003

\REF Sha Shastri \Jref
    B.S. Shastry
    "Taking the square root of the discrete $1/r^2$ model"
    Phys.Rev.Lett. @69(1992)164--167

\REF Sla Slawny  \Gref
    J. Slawny
    "Low temperature proerties of classical lattice systems"
    \inPr C. Domb, J.L. Lebowitz
    "Phase transitions and critical phenomena, vol.11"
    Acad.Press, New York 1987

\REF {St\o} Stormer \Jref
    E. St\o rmer
    "Symmetric states of infinite tensor products of C* -algebras"
    J.Funct.Anal. @3(1969) 48--68

\REF We1 Delhi \Gref
    \sameauthor. {}
   "Large Deviations and mean-field quantum systems"
   \inPr L. Accardi
   "Quantum probability and related topics, vol. VII"
   World Scientific, Singapore 1992

\REF We2 FCL  \Gref
    \sameauthor. {}
    "Finitely correlated pure states"
    \inPr  M. Fannes, C. Maes, A. Verbeure
    "On three levels; micro-, meso-, macro-approaches in physics"
    Plenum Press, New York 1994

\REF We3 KAC \Gref
    \sameauthor. {}
    "Quantum lattice systems with infinite range interactions"
    In preparation

\REF We4 CLQ \Gref
    \sameauthor. {}
    "The classical limit of quantum theory"
    Preprint Osnabr\"uck, RFW-15Nov94

\REF Wo1 WoRims  \Jref
    S.L. Woronowicz
    "Twisted SU(2) group. An example of a non-commutative
    differential calculus"
    Publ.RIMS, Kyoto @23(1987) 117--181

\REF Wo2 WoCMP \Jref
    S.L. Woronowicz
    "Compact matrix pseudogroups"
    Commun.Math.Phys. @111(1987) 613--665

\REF YY YY \Jref
    C.N. Yang, C.P. Yang
    "One-dimensional chain of anisotropic spin-spin interactions.
     I. Proof of Bethe's hypothesis for ground state in a finite
    system"
    Phys.Rev. @150(1966)321--327;
    \more
    ``II. Properties of the ground-state energy per lattice site for
    an infinite system'',
    \Jn Phys.Rev. @150(1966)327--339;
    \more
    ``III. Applications'', \Jn Phys.Rev. @151(1966)258--264

\bye